# Analytical capability of in-situ K–Ar isochron dating on Mars: assessment from mineral compositions of Martian meteorites

Hikaru Hyuga[1], Yuichiro Cho[1], Yayoi N. Miura[2], Takashi Mikouchi[3], and Seiji Sugita[1]

1. Department of Earth and Planetary Science, University of Tokyo, 7-3-1 Hongo, Bunkyo, Tokyo, 113-0033 Japan
2. Earthquake Research Institute, University of Tokyo, 1-1-1 Yayoi, Bunkyo, Tokyo, 113-0032, Japan
3. The University Museum, University of Tokyo, 7-3-1 Hongo, Bunkyo, Tokyo, 113-0033 Japan

e-mail: hyuga@eps.s.u-tokyo.ac.jp

## Abstract

Dating rocks with a precision of ±200 Myr (2σ) has been required to understand the Martian habitability and volcanic history around 4000 Myr ago. In situ potassium-argon (K-Ar) dating techniques employing spot-by-spot laser ablation have been developed for isochron dating on Mars. The precision of isochron ages is predominantly determined by the relationship between the laser spot diameter and the mineral grain size. However, the achievable age precisions for the realistic mineralogy of Martian rocks were not investigated yet. This study simulates isochrons under various conditions, including laser spot size, measurement errors for K and Ar, the number of isochron points, and the mineral compositions of representative Martian meteorites (NWA 817, Zagami, and NWA 1068) analyzed with an electron microprobe. We find that attaining the 200 Myr precision necessitates an isochron data range wider than 6, a laser spot diameter of 250 μm, and measurement errors of 10% for both K and Ar. Furthermore, minimizing the variance in measurement errors between K and Ar is essential for increasing age accuracy. This investigation demonstrates that the required precision for Mars missions is achievable with realistic instrument settings, thus supporting the viability of in situ K-Ar geochronology on Mars.

## INTRODUCTION

Accurate age determination is essential for understanding the history of planetary bodies. Establishing Martian chronology is particularly important for elucidating its past potentially habitable climate (Wordsworth, 2016; Ehlmann et al., 2016). Absolute dating with a precision of 200 Myr would narrow down the timing of the cessation of fluvial activity and potential habitability in warm and wet climates (Cohen et al., 2021). Also, understanding the age and composition of igneous rocks generated through volcanic activities is useful for reconstructing the evolution of the geology of Mars (Doran et al., 2004; Cohen et al., 2021; Mouginis-Mark et al., 2022). In fact, various strategic plans, including the Decadal Strategy 2023–2033 by the National Academies of Sciences, Engineering, and Medicine, have emphasized the importance of determining the absolute age of Mars (National Academies of Sciences, Engineering, and Medicine, 2023).

The age of the Martian surface has been estimated by crater counting. The crater production function for Mars is derived by applying a correction to the lunar crater production function (Hartmann & Neukum, 2001). This approach, however, introduces significant uncertainties, with estimates suggesting a potential error of up to 2 billion years in the absolute age of the Martian surface (Doran et al., 2004). To reduce the large uncertainties in Martian chronology, samples with both known formation ages and locations are essential. While NASA plans sample return missions from Mars (Beaty et al., 2019), in situ dating is still beneficial for establishing a robust chronology that requires absolute age measurements across varied geologic units on Mars. Such measurements would need multiple one-way missions, which should be less costly and more frequent than sample return missions.

In this context, various in situ dating techniques have been developed, including Rb–Sr, K–Ar, and Pb–Pb methods (Anderson et al., 2019; Cohen et al., 2014; Cho et al., 2016; Anderson et al., 2017). Here we focus on the K–Ar method. The K–Ar method quantifies the ratio of $^{40}$K to its radioactive decay product $^{40}$Ar in rock samples. It is also important to note that K is one of the major elements, typically more abundant than Rb, Sr, or Pb. Remote sensing data from the gamma ray spectrometer (GRS) have revealed relatively low K concentrations ranging from 2000–6000 ppm in Martian surface (Boynton et al., 2007). Most Martian meteorites also show K concentrations below 3000 ppm (Tian et al., 2021). Data from Curiosity's alpha particle X-ray spectrometer (APXS) identified rocks with elevated bulk K concentrations, 2–8 times higher than the average Martian crust, but most measurements fall within a few thousand ppm (Thompson et al., 2016).

While the Curiosity rover conducted pioneering in situ K–Ar dating experiments on Mars (Farley et al., 2014), the measurement results yielded approximately a factor two of uncertainty. Specifically, the K–Ar age measurements for the authigenic mineral jarosite varied from 2.12 Ga to 4.07 Ga, mainly due to challenges in quantifying the amount of Ar lost from an amorphous/phyllosilicate fraction before the measurements (Martin et al., 2017). Furthermore, Vasconcelos et al. (2016) pointed out that the oven used for Ar extraction may have not sufficiently released Ar trapped within the crystal lattice. Moreover, K and Ar were measured in separate aliquots, preventing their measurement in exactly the same sample.

New dating instruments are being developed to address these challenges (e.g., Cohen et al., 2014; Cho et al., 2016; Devismes et al., 2016; Cattani et al., 2019; Solé 2021). This method integrates laser-induced breakdown spectroscopy (LIBS) and a mass spectrometer (MS) to conduct laser spot analysis of rocks. This method has the advantage of utilizing a combination of equipment and techniques already used in planetary missions, such as LIBS and a quadrupole mass spectrometer (QMS). This approach enables isochron analyses by applying a pulsed laser with a spot diameter of several hundred micrometers to the sample. The method presents multiple advantages, such as reducing age determination uncertainties through isochron methods and possibility of evaluating the effects of trapped Ar. It also enables the measurement of rocks with high Ar retention through complete vaporization by laser ablation. Previous studies have employed this technique to analyze K-rich terrestrial rocks and basalts (Cho & Cohen, 2018; Cho et al., 2016).

The effectiveness of isochrons, however, depends on several factors. The precision of an isochron slope, which equates to the dating precision, is influenced by individual data point accuracy and the data's range. If the data points cluster around a similar K concentration, one cannot obtain precise slope values. In contrast, a broader data range enhances the precision. To quantitatively assess these factors, Bogard (2009) examined the impact of measurement range on the precision of K–Ar isochron slopes. Assuming bulk measurements of rocks with identical formation ages, the study generated hypothetical isochron data points and analyzed the resultant isochron slope error. Findings indicated that with a data range exceeding a factor of 2 and assuming a 10% measurement accuracy for K and Ar concentrations within a ten-point dataset, the slope error is approximately several tens of percent. Bogard (2009) then pointed out that in situ K–Ar isochron dating is challenging because finding rocks exhibiting such a wide bulk K concentration seemed unrealistic.

The laser spot analysis approach can overcome this limitation by producing internal isochrons for different minerals of a single rock. The data range attainable with this method depends on factors such as the diameter of laser spot, the mineral size distribution of the rock, and the K concentration in each mineral. A smaller laser spot broadens the K concentration range on the isochron but reduces the sample volume, which affects extracting Ar amount and measurement accuracy. Conversely, a larger laser diameter complicates the isolation of individual minerals for measurement, making all data points close to the bulk K concentration of the rock. Thus, a trade-off exists between K concentration range and amount of extractable Ar, factors governed by laser diameter and mineral lithology. Given that lithology depends on Martian rock, the controllable parameter is the laser diameter. Once lithology is determined, the optimal laser diameter can be determined, thereby determining the dating precision.

However, no studies have investigated isochrons obtained through laser spot analysis while taking into account the lithology and formation ages of actual Martian rocks. Without knowledge of the mineral size distribution and K concentration in Martian rocks, it is challenging to ascertain the optimal instrument conditions for in situ dating experiments. Determining factors such as the ideal laser diameter and the required measurement accuracy for both K and Ar is crucial to effectively address the scientific questions associated with Martian geochronology. Thus, in this study, we evaluate the dating precision of Martian meteorites by creating simulated isochrons. These simulations are based on the mineralogical composition and K concentration distributions in the meteorites. We aim to determine the laser diameter and measurement accuracy for both K and Ar concentrations required to achieve a dating precision of 200 Myr, as outlined in the NASA Technology Roadmaps (2015).

This paper is structured as follows: the Methods section describes samples, the analytical method applied to Martian meteorites, and the methodology for simulating isochrons using the obtained data; the Results section presents the results of an electron probe microanalyzer (EPMA), including K concentration maps and the K concentration distribution derived from laser spot analysis using these maps. In the Discussion section, we discuss the optimal instrumental conditions required to achieve an age determination with the precision of 200 Myr based on the simulated isochrons.

## METHODS

The precision of isochron dating depends on the range of data points displayed on the isochron plot and the error of each data point. To accurately simulate the capability of isochron for K–Ar dating, the spatial distribution of K (i.e. mineralogy) in actual Martian meteorites should be investigated.

## Samples

We analyzed three Martian meteorites of different classification and characteristics: NWA 817, Zagami, and NWA 1068 (Fig. 1).

NWA817, a nakhlite, comprises of approximately 69% augite, 10% olivine, 20% mesostasis, and 1% Fe–Ti oxide (vol%) (Sautter et al., 2002). Its K–Ar age is estimated at 1.3 Ga (Mathew et al., 2003), but the Ar–Ar age remains unreported.

Zagami is classified as a basaltic shergottite containing 74–80% pyroxene, 10–20% maskelynite, 2–4% mesostasis, 2% Fe–Ti oxides, phosphate, and shock melt (McCoy et al., 1992; Borg et al., 2005). The Ar–Ar age ranges from 200–250 Ma (Bogard & Park, LPSC, 2006) and its Rb–Sr age is 166 ± 6 Ma (Borg et al., 2005). The Ar–Ar ages appear older due to trapped $^{40}$Ar during crystallization and non-radiogenic excess $^{40}$Ar (Bogard & Garrison, 1999; Bogard & Park, 2008).

NWA1068, a picritic shergottite, primarily consists of 52% pyroxene, 22% maskelynite, 21% olivine, 2% phosphate, 2% opaque oxides, and 1% K-rich mesostasis (Barrat et al., 2002). It contains fragments of olivine-rich lithology and other facies within a basaltic shergottite (Barrat et al., 2002; Goodrich, 2002). Its cosmic-ray exposure age is approximately 2 Ma, similar to other shergottites (Mathew et al., 2003). The K–Ar and Sm–Nd ages are 610 Ma (Mathew et al., 2003) and 185 ± 11Ma (Shih et al., 2003), respectively.

(a) NWA 817

(b) Zagami

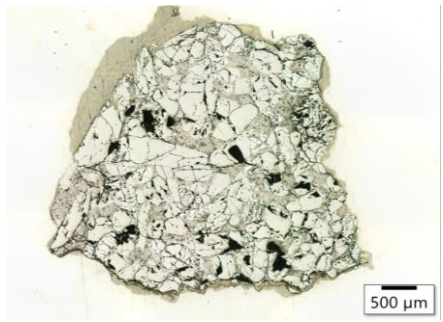

(c) NWA 1068

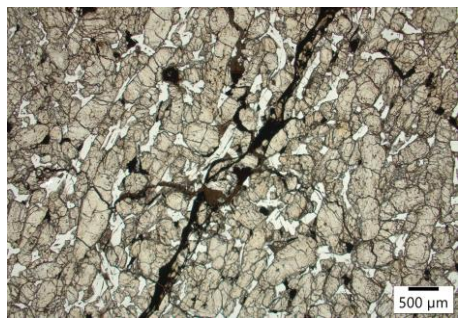

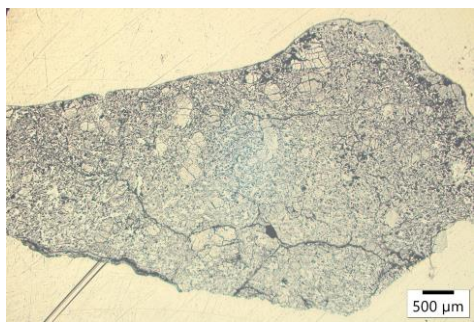

**Figure 1**. Samples prepared for our EPMA analysis. (a) NWA 817 (b) Zagami (c) NWA 1068.

## Electron probe micro analyzer setting

Elemental mapping and quantitative analysis were conducted on polished thin sections of the meteorite samples using an EPMA (JEOL JXA-8900L). Operating conditions included an accelerating voltage of 15 kV, beam currents of $8 \times 10^{-8}$ A for mapping and $8 \times 10^{-9}$ A for quantitative analysis. Spot sizes varied, with 5 μm in diameter for NWA 817 and NWA 1068, and 7 μm for Zagami. The analyses covered elements such as Na, Mg, K, Fe, Ti, Al, Si, Ca, Mn, and S for mapping, while the quantitative analysis focused on all but S. The EPMA was calibrated using established natural and synthetic standards. Mineral identification was based on stoichiometry. Each mineral section underwent analysis at 6 to 18 points, and the glass portion at 10 points, to assess the K concentration variations within each phase.

The mapping area covered 2.5 mm × 2.5 mm for NWA 817, 6.5 mm × 6.5 mm for Zagami, and 3.0 mm × 3.0 mm for NWA 1068. These sizes are the largest square area available within the prepared meteorite thin sections, exceeding the sizes of mineral crystals and thereby ensuring representative mineral composition (discussed further in K concentration map section of results).

## K concentration map and histogram

From the elemental maps of Fe, Al, Ca, and K, distinct areas for each mineral type were extracted. The K concentration maps of each meteorite were created by assigning the average K concentration of each mineral measured with the quantitative analyses. To predict K concentrations under different laser spot diameters, histograms were derived from these maps. In previous studies that employed laser ablation, several hundred to a thousand of laser pulses were applied to the same spot to enhance the signal-to-noise ratio in LIBS and to increase the amount of $^{40}$Ar extracted from the rock samples (Cohen et al. 2014; Cho et al. 2016). Ablation craters were several hundred μm deep, and the averaged K concentration within the crater were measured with LIBS.

The elemental maps do not reveal the mineral distribution in depth direction, which raises challenges for using 2-D surface data to infer 3-D characteristics. For example, extrapolating a uniform concentration depth from a 2-D map is problematic when the excavated depth by laser (e.g., 1000 μm) is significantly larger than the mineral size, such as 100 μm. To address this limitation, we averaged the 2-D K map in the vertical direction (y-direction), assuming that the distribution to the y-direction similarly extends to the depth direction (z-direction). Then we produced a depth-averaged K map (Fig. 2).

By calculating the average K concentration using the depth-averaged K map, we obtained the average potassium concentration taking the approximate depth distribution into account. We computed the average K concentration in circular regions on this map since we assumed cylindrical ablation craters with equivalent depth and diameter. A moving circular region of interest (ROI), simulating different laser spot sizes, scanned the map to generate a probability distribution of K concentrations. This assumed random laser targeting across the meteorite surface.

In this procedure, the top-left pixel of the K map serves as the coordinate origin (0 pix, 0 pix), with the +x and +y directions extending to the right and bottom, respectively. The algorithm averages values along rows parallel to the +y direction, equal in length to the laser spot depth, for each pixel in the map. This creates a depth-aware map. The algorithm then calculates the average K concentration within the ROI, starting from the top-left corner of the map. The ROI's center coordinates shift incrementally in the +x and +y directions, repeating the calculation at each new position to cover the entire map.

To provide practical insights for actual laser ablation experiments, we considered laser spot diameters of 50, 250, 500, 700 and 1000 μm. Histograms of the K concentrations were generated for each meteorite and laser diameter, with a bin width set at 0.05 wt%.

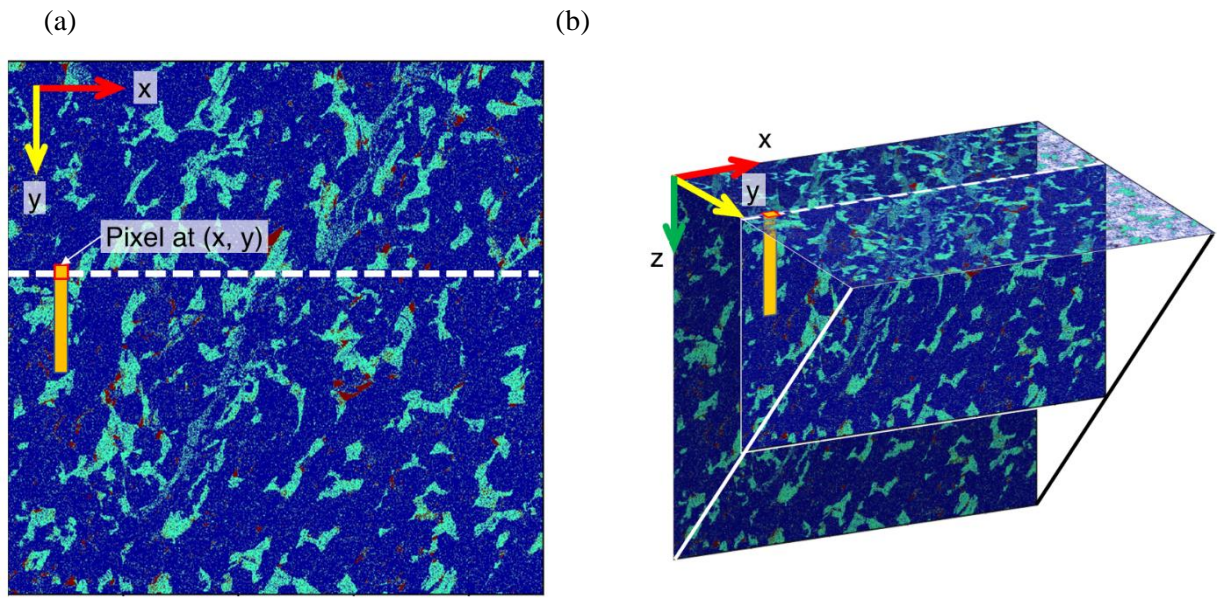

**Figure 2.** Concept of depth integration using a K concentration map. (a) conceptual diagram illustrating depth integration in 3-D space on a 2-D map. When calculating the depth-integrated K value for a pixel on the white dashed line (red box), the averaged K value in the orange area is allocated to the pixel in the red box, assuming that the orange area continues in the depth direction, i.e., +y-direction in the 2-D map. (b) 3-D representation of this process.

## Volume of extracted Ar

We estimated the accumulation of radiogenic $^{40}$Ar in rocks based on given K distributions and various ages:

10 Ma, 100 Ma, 300 Ma, 500 Ma, 1 Ga, 1.5 Ga, 2 Ga, 2.5 Ga, 3 Ga, 3.5 Ga, 3.8 Ga, 4 Ga, 4.2 Ga, and 4.5 Ga. Specifically, we employed an age of 4000 Ma in multiple parameter studies, given that the ancient warm and wet climate on Mars is thought to have existed between 3.5 and 4.0 Ga (Ehlmann et al., 2016).

The $^{40}$Ar concentration at each measurement spot was derived from the average K concentration within the laser spot and the assumed rock formation age. The amount of $^{40}$Ar, to be measured with a mass spectrometer, was calculated by multiplying the $^{40}$Ar concentration by the volume of the sample ablated by the laser. The calculations involved the following equations.

$$t = \frac{1}{\lambda} \ln\left(1 + \frac{C \alpha M_{40Ar}}{\rho V [K_2O]}\right) \quad (1)$$

$$C = \frac{\lambda M_K}{\lambda_e \left(\frac{2M_K}{M_{K2O}}\right) F_{40K} M_V} \times 100 \quad (2)$$

where $\alpha M_{40Ar}$ (cm$^3$STP), [K$_2$O] (wt%) represents the absolute $^{40}$Ar amount and K$_2$O concentration, respectively; $\rho$ is the density of the rock, and $V$ is the volume of the ablated sample. A density of basalt $\rho$ = 2.7 g/cm$^3$ was assumed for our calculations. $F_{40K}$ = 0.01167% denotes the natural fraction of $^{40}$K in total K. $2M_K/M_{K2O}$ = 0.830 where $M_K$ is the atomic weight of K and $M_{K2O}$ is the molecular weight of K$_2$O. $M_V$ = 22400 cm$^3$STP/mol signifies the molar volume of a gas at standard temperature and pressure. $\lambda$ = 5.54×10$^{-10}$ [yr$^{-1}$] and $\lambda_e$ = 5.81×10$^{-11}$ [yr$^{-1}$] refer to the total and electron-capture decay constants of $^{40}$K, respectively (Steiger and Jäger, 1977).

## Isochron simulation

We generated simulated isochrons by varying parameters such as K distributions corresponding to different meteorites, assumed formation ages, laser spot diameters, numbers of laser spots, and measurement errors in K or $^{40}$Ar concentrations. The formation age for these simulations ranged from 10 to 4500 Ma, and the number of spots on each isochron varied from 5 to 50 in increments of 5. We also accounted for errors in K and $^{40}$Ar concentrations, which correspond to the 1σ value of the normal distribution and define the size of the error bars when variations are applied to these concentrations. The following steps outline the procedure for generating a hypothetical isochrons with $N$ data points:

1. K concentration values were randomly generated based on the probability distribution obtained from the histogram (K concentration map and histogram section). The range of K concentration, a parameter indicative of the data range of an isochron, is defined as K conc. range = K$_{max}$/K$_{min}$ (3)

2. The concentration of radiogenic $^{40}$Ar was calculated using Eq. (1), considering the K concentration and the assumed formation age. Non-radiogenic $^{40}$Ar from the atmosphere or magma (Bogard et al., 2009) was not considered. Ideally, its contribution can be evaluated as the intercept of an isochron.

3. For each K–$^{40}$Ar data point, deviations were introduced as random numbers generated from a normal distribution. The 1σ of this distribution was defined as the "K and Ar measurement errors" parameter, which varies from 0 to 20% in this study.

By applying these steps, we simulate realistic isochron analyses for Martian rocks with various mineral compositions, assumed ages, laser spot sizes, and measurement errors.

We employed the method by York (2004) to derive isochron slopes and intercepts. This method uses the inverse square of the measurement errors as a weight for each data point. For this purpose, we developed a Python-based program and validated it against results from previous studies (Cho et al., 2016; Cho & Cohen, 2018). Ages were determined from the isochron slopes using Eq. (4), while the precision of the ages was assessed using Eq. (5), which is a differential form of Eq. (4) that accounts for error propagation. *ΔSlope* represents the standard deviation of the fitted isochron slope.

$$t = \frac{1}{\lambda}\ln(1 + C \times Slope) \qquad (4)$$

$$t_{error} = \frac{C \times \Delta Slope}{\lambda \times (1 + C \times Slope)} \qquad (5)$$

## RESULTS

This section evaluates isochron dating precision under various conditions. First, elemental maps of the three Martian meteorites based on EPMA data are presented. Subsequently, the types of minerals in each meteorite and K concentration maps derived from the EPMA results are displayed. K concentration histograms and Ar extraction volumes for the three Martian meteorites, considering different laser spot diameters, are then provided. The subsequent section details the quantitative evaluation of factors such as mineral composition (meteorite type), laser spot diameter, and the number of isochron data points on a precision of isochron. Statistical analysis is performed to assess the precision of isochron, considering laser spot diameter, the number of isochron data points, formation age, and the precision of K and Ar concentration measurements as parameters. Finally, we explore trends in the accuracy of age determination by changing these parameters.

## EPMA mapping

The mineral compositions and locations of K-rich phases in each meteorite were determined through the EPMA elemental maps. Our results were consistent with previous studies of the same three meteorites (Barrat et al., 2002; Borg et al., 2005; Mathew et al., 2003, McCoy et al., 1992). Specifically, NWA 817 showed pyroxene-dominated texture with K-rich mesostasis; Zagami exhibited a pyroxene-rich mineralogy with relatively K-rich maskelynite and even more K-rich mesostasis of approximately 100 μm in size; NWA 1068 showed olivine-rich mineralogy with K-rich mesostasis (Fig. 3).

The mineral sizes in the meteorites showed considerable variation. The largest phenocryst of olivine observed in NWA 817 was approximately 500 μm in diameter. Some crystals even reached lengths of approximately 1 mm. Pyroxene crystals in Zagami have sizes of approximately 800 μm. Our observations of NWA 817 and Zagami validate that the modal mineral abundances, as calculated from area ratios (Table 1), are consistent with the results of previous studies (Sautter et al., 2002; McCoy et al., 1992). In NWA 1068, olivine exhibits the largest crystal size, with a maximum size of approximately 300 μm. The other phases have sizes smaller than 100 μm and exhibit indistinct shapes. In the analyzed area of this study, the modal abundance except for olivine in NWA 1068 was consistent with the reported values within 2 vol%. The modal abundance of olivine measured in this study was 13 % while Barrat et al. (2002) reported 21%. Given the presence of mega olivine crystals up to 2 mm in this meteorite (Barrat et al., 2002). This difference could be due to the inherent heterogeneity of the meteorite. Thus, the areas measured in this study represent an ordinary region within each meteorite.

(a) NWA 817

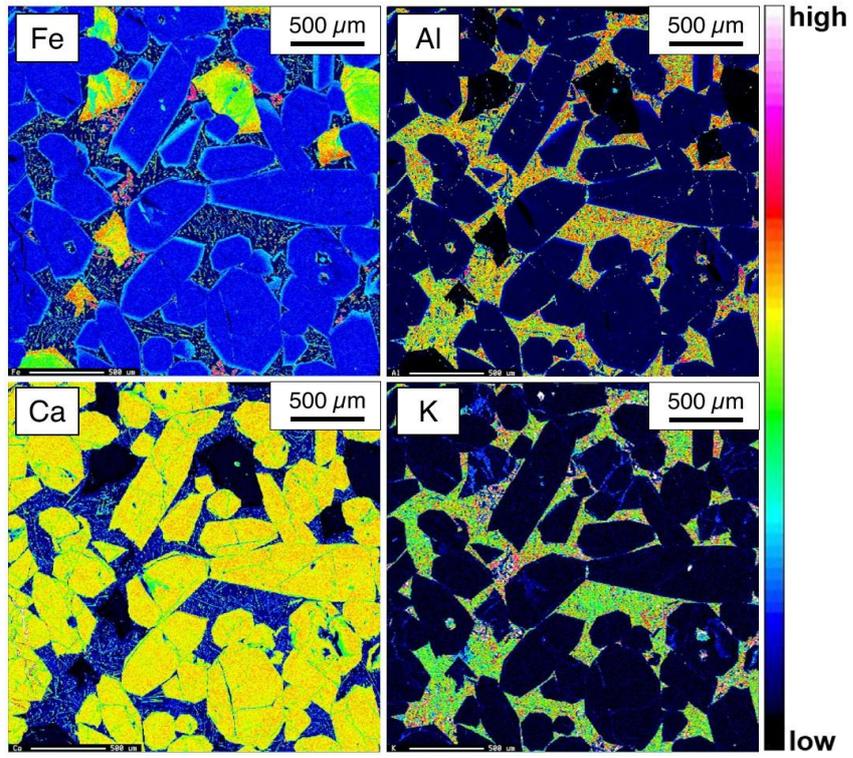

(b) Zagami

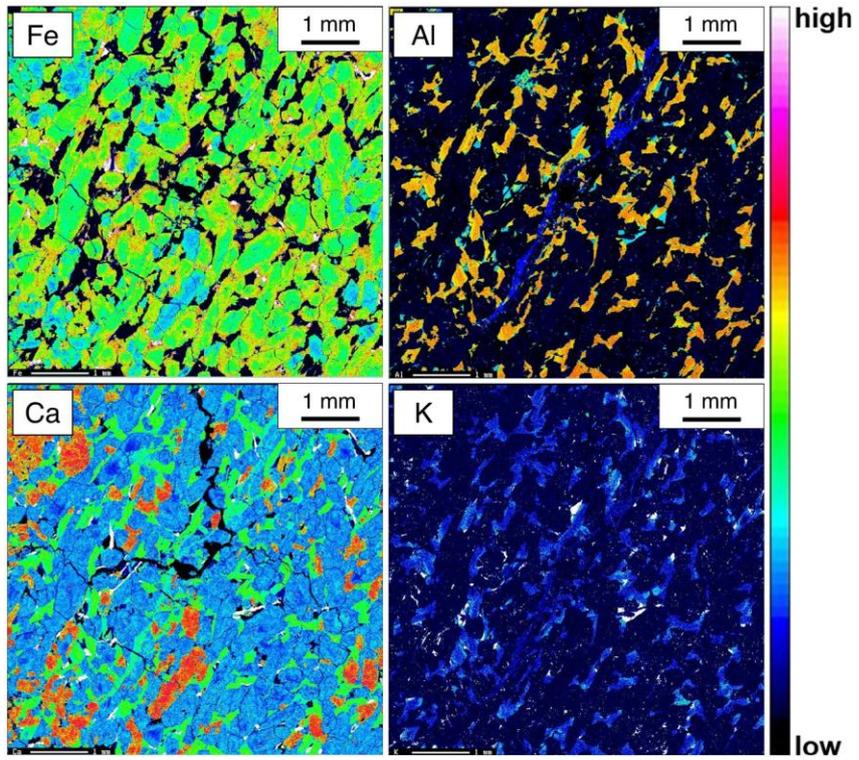

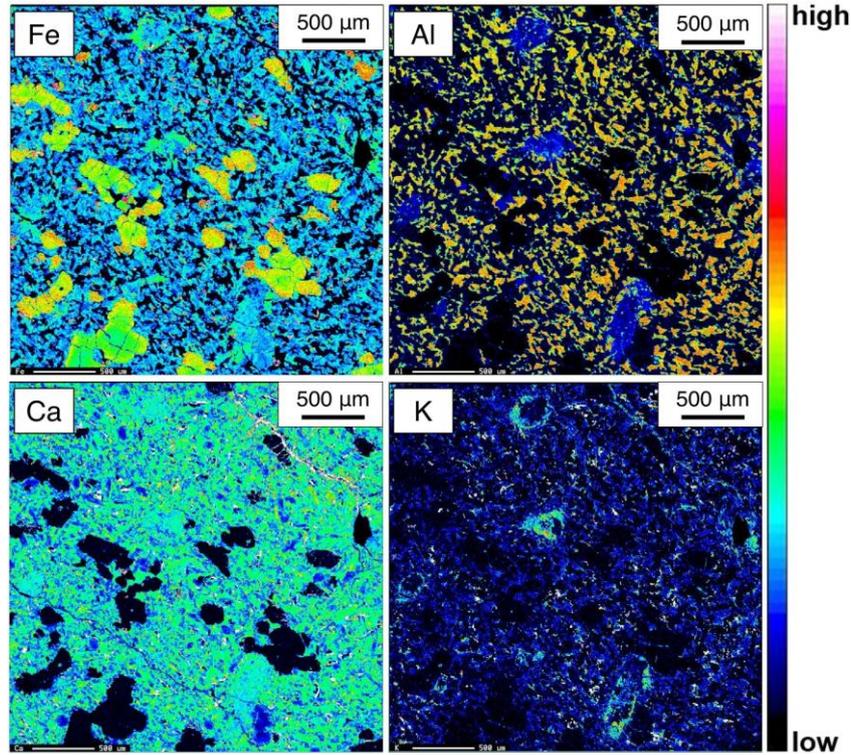

**Figure 3.** Elemental maps of (a) NWA 817, (b) Zagami, and (c) NWA 1068.

**Table 1.** Average K concentration in each mineral phase.

| Sample | Mineral | $K_2O$ [wt%] | Modal abundance [vol%] |
| --- | --- | --- | --- |
| NWA817 | Olivine | 0.00 ± 0.01 | 8.8 |
|  | Pyroxene | 0.01 ± 0.01 | 68 |
|  | Fe-Ti oxides | 0.03 ± 0.02 | 0.7 |
|  | Mesostasis | 2.05 ± 1.25 | 22 |
| Zagami | Pyroxene | 0.01 ± 0.01 | 74 |
|  | Maskelynite | 0.42 ± 0.16 | 19 |
|  | K-rich mesostasis | 4.83 ± 2.94 | 2.7 |
|  | Others | 0.00* | 4.9 |
| NWA1068 | Olivine | 0.01 ± 0.01 | 13 |
|  | Pyroxene | 0.04 ± 0.03 | 54 |
|  | Maskelynite | 0.36 ± 0.07 | 24 |
|  | K-rich mesostasis | 5.25 ± 1.43 | 2.2 |
|  | Shock melt | 0.50 ± 0.33 | 5.7 |
|  | Others | 0.00* | 1.7 |

* "Others" include Fe–Ti oxides and calcium phosphates, of which K concentrations were assumed to be 0.00 wt%.

# K concentration map

Potassium concentration maps for each meteorite were derived based on the quantitative elemental analysis. Table 1 presents the average K concentrations for each mineral. In NWA 817, K is primarily concentrated in the mesostasis between minerals, yielding approximately 2.05 wt% $K_2O$. Zagami reveals high K concentrations in both maskelynite and mesostasis, at 0.42 wt% and 4.83 wt%, respectively. For other lithologies such as shock melt, calcium phosphate, and Fe–Ti oxides, a K concentration of 0.0 wt% was assumed based on the mapping data. In NWA 1068, K concentrations were 0.36 wt% in maskelynite, 5.25 wt% in K-rich mesostasis, 0.50 wt% in other components such as shock melt (Fig. 3c), and 0.00 wt% in calcium phosphate and Fe–Ti oxides. The bulk K concentrations for NWA 817, Zagami, and NWA 1068 were reported as 0.42, 0.14, and 0.20 wt% $K_2O$ respectively (Udry and Day, 2018; Lodders, 1998; Filiberto et al., 2010), while our EPMA analysis yielded 0.47, 0.22, and 0.25 wt%, respectively. This apparent deviation can be explained by the heterogeneity of K concentrations within K-rich mesostasis. The relatively large standard deviation in mesostasis measurements can be attributed to intrinsic variations within the mesostasis. While these fluctuations are detectable at the 5 μm resolution of the EPMA, they are averaged out over the larger laser spot diameter (> 50 μm) and therefore do not significantly influence subsequent analyses.

Figures 4(a, c, e) display K concentration maps with corresponding quantitative data for each mineral. The sizes of K-concentrated areas in NWA 817, Zagami, and NWA 1068 were approximately 500 μm, 100 μm, and 10 μm, respectively. Variations in mineral sizes across these meteorites may reflect differences in the cooling rates of their parent magmas, which could be influenced by their formation depths.

Figures 4(b, d, f) show the depth-averaged K concentration maps as detailed in the Methods section. In these maps, the concentration represented in each pixel at coordinates $(x, y)$ averages the K concentration from $(x, y)$ to $(x, y + y_d)$, where $y_d$ represents the size in the depth direction to be integrated. For illustrative purposes, these figures show cases for a depth ($y_d$) of 500 μm. Note that for the 2.5 × 2.5 mm chip of NWA 817, the bottommost 0.5 mm pixels are excluded from Fig. 4 (b) because depth-averaging was not feasible. In contrast, the larger chips of Zagami and NWA 1068 include data of the areas.

The spatial distribution of K-rich phases in these depth-averaged maps correlates with mineral grain size. For instance, Zagami's crystals were larger than 500 μm, and areas of elevated K concentration remain visible after averaging (Fig. 4(d)). Contrary, NWA 1068, which has smaller crystals at a scale of approximately 100 μm, exhibits a more uniform K concentration in its depth-averaged map (Fig. 4(f)). In NWA 817, regions of higher K concentration are also observed, resembling Zagami due to its coarser grain structure (Fig. 4(b)). Thus, K concentrations measured with LIBS are subject to variations based on the spatial scale of minerals and the size of the laser spot employed.

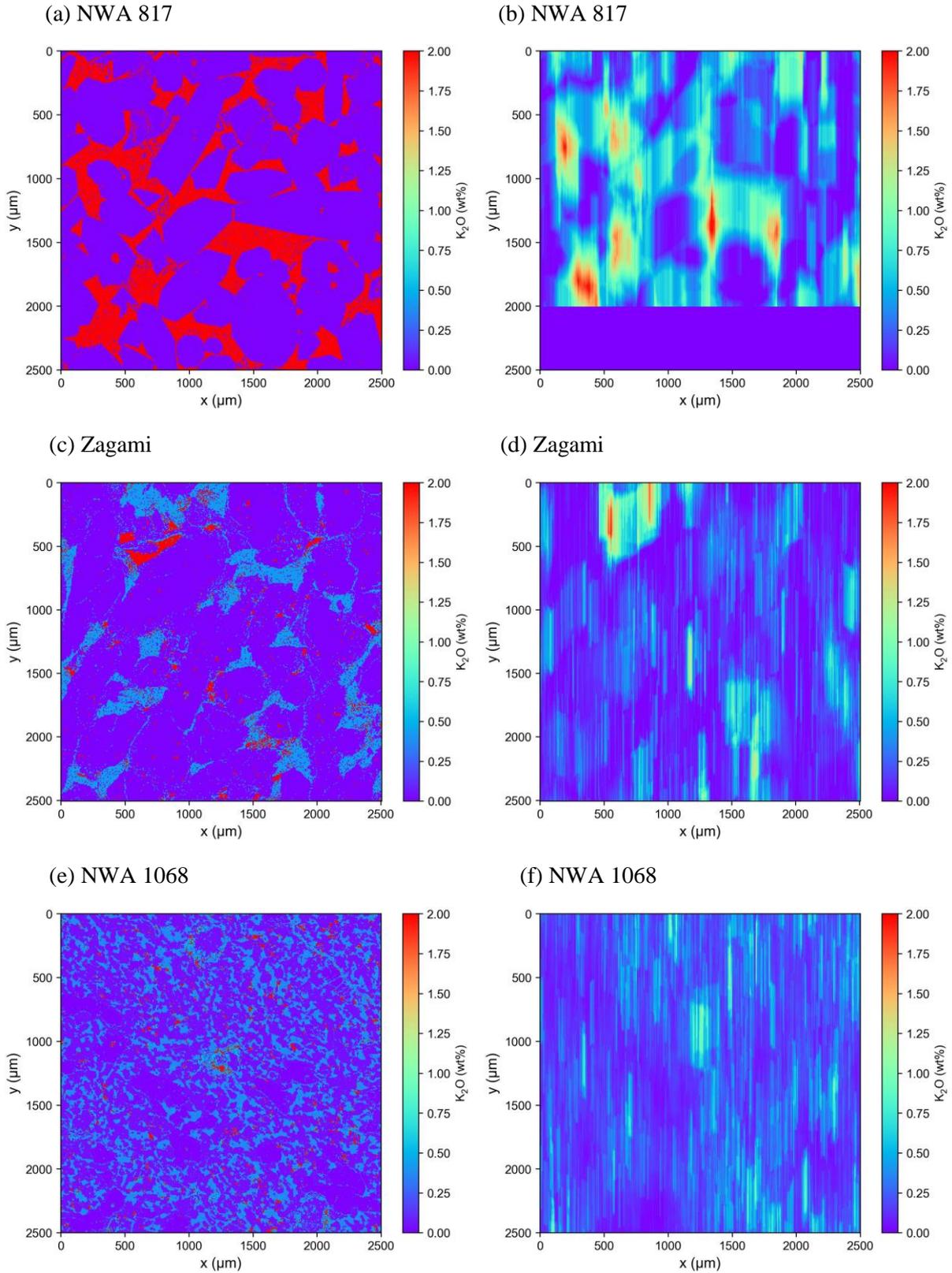

**Figure 4.** K concentration maps (a, c, e) and depth-averaged K concentration maps (b, d, f). (a, b) NWA 817 (c, d) Zagami, (e, f) NWA 1068. The depth-averaged maps (d, e, f) assume an ablated depth of 500 μm. The images are 2.5 × 2.5 mm, and the areas for Zagami and NWA 1068 are cropped sections from the broader analysis region.

## Variability in K concentration across meteorites and laser spot diameters

The attainable range of K concentrations varies according to both laser spot diameter and mineral distribution (Figs. 4b, d, f). Figure 5 displays histograms of $K_2O$ concentrations (wt%) achievable at different laser spot diameters for the three meteorites. Decreasing laser diameter expands the K concentration range, while increasing it narrows the range as concentrations are spatially averaged toward the bulk K concentrations. Specifically, NWA 817 exhibited $K_2O$ concentrations ranging from 0.33–0.63 wt% at a 1000 μm laser spot, 0.14–1.06 wt% at 500 μm, and 0.01–1.77 wt% at 250 μm. Similarly, Zagami showed ranges of 0.10–0.46 wt% at 1000 μm, 0.04–0.92 wt% at 500 μm, and 0.01–2.18 wt% at 250 μm. NWA 1068 exhibited ranges of 0.18–0.30 wt% at 1000 μm, 0.11–0.42 wt% at 500 μm, and 0.02–0.81 wt% at 250 μm. The fine-grained NWA 1068 yielded a narrower range of K concentrations for each laser spot diameter.

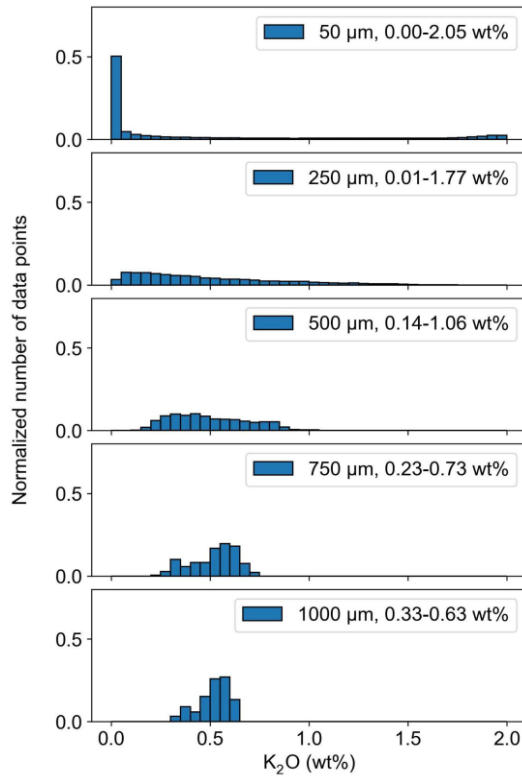
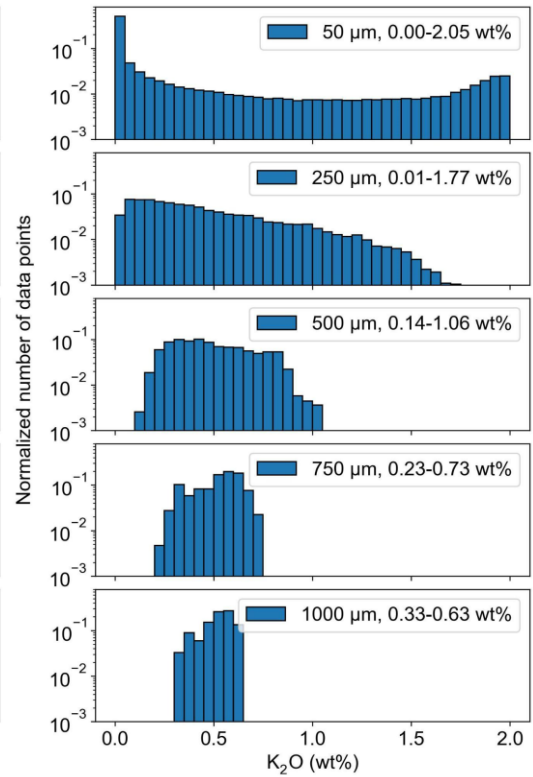

(a) NWA 817 (linear)  (b) NWA 817 (log)

(c) Zagami (linear)

(d) Zagami (log)

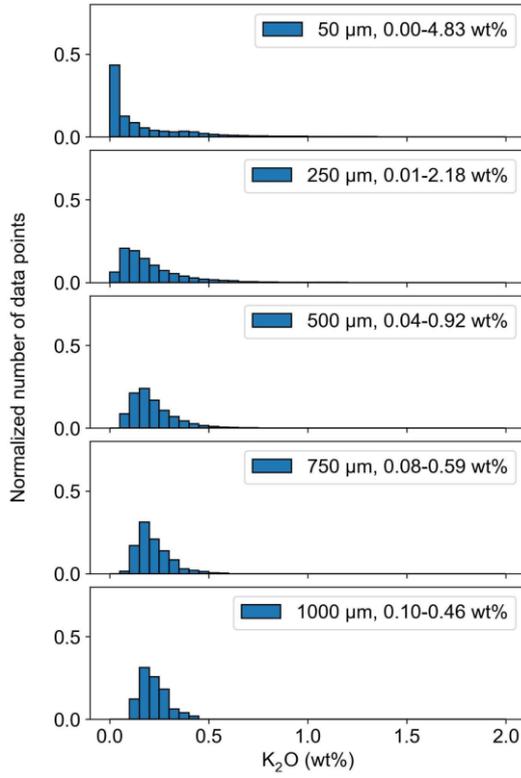
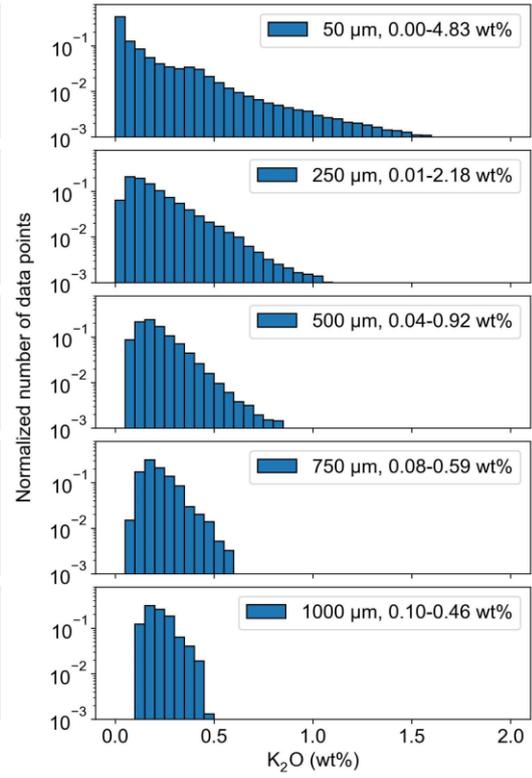

(e) NWA 1068 (linear)

(f) NWA 1068 (log)

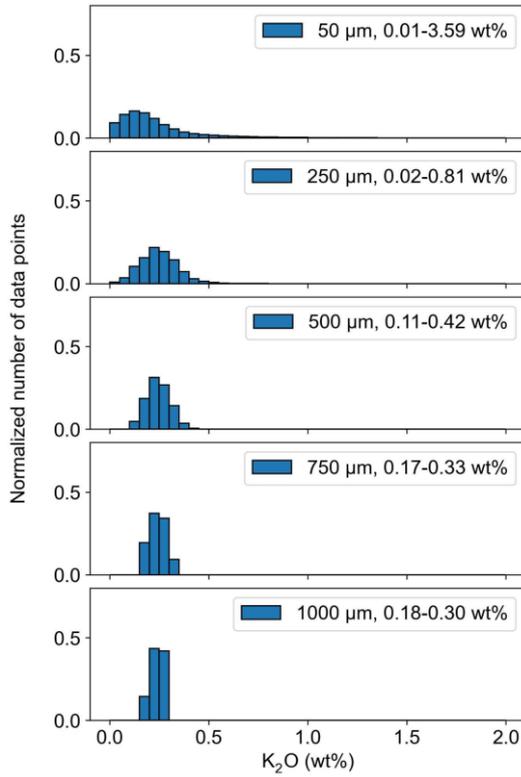
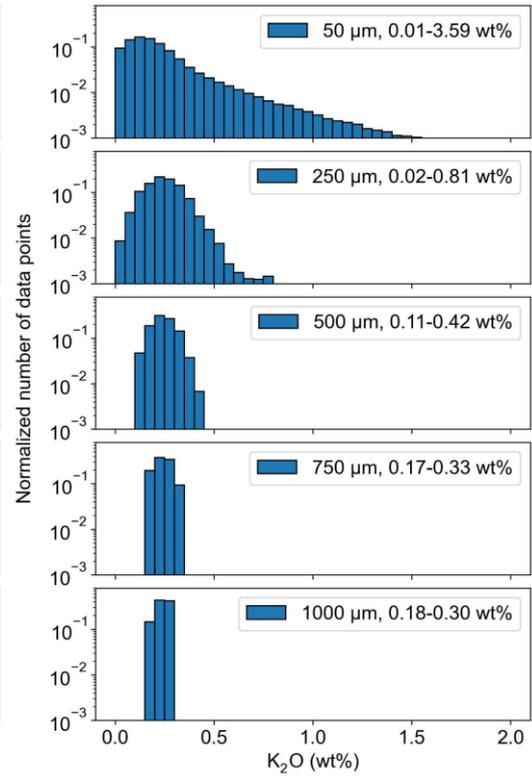

**Figure 5.** Histogram of K concentrations. (left) linear plot (right) logarithmic plot. (a, b) NWA 817, (c, d) Zagami, (e, f) NWA 1068.

# Laser spot size and amount of Ar extraction

This section investigates the absolute quantity of Ar liberated from the meteoritic samples via laser ablation. The analysis employs assumed age and K concentration maps obtained in the previous section. The extracted Ar amount correlates directly with the ablated material volume. Figure 6 illustrates the distribution of K concentration and $^{40}$Ar content for different laser spot sizes and meteorites, assuming a sample age of 4000 Ma. The data are presented as violin plots, where the plot width represents the data distribution for a given spot size.

For example, using a 250 μm spot diameter, the NWA 817 sample most likely exhibits $K_2O$ concentration of approximately 0.1 wt%, with a potential range from 0–1.7 wt%. Accordingly, the $^{40}$Ar content peaks around $3\times10^{-13}$ mol, spanning a range from $2\times10^{-14}$ to $2\times10^{-12}$ mol. Red and blue dashed lines in Fig. 6 indicate the quadrupole mass spectrometer (QMS) background and K quantification limit with LIBS as established by Cho and Cohen (2018) (0.1 wt% K concentration and $9\times10^{-16}$ mol $^{40}$Ar content).

At a 50 μm laser spot size, the anticipated $^{40}$Ar content ranges for NWA 817, Zagami, NWA 1068 were $0.00 - 1\times10^{-13}$ mol, $0.00 - 2\times10^{-13}$ mol, and $1\times10^{-15} - 2\times10^{-13}$ mol, respectively. At 250 μm, these ranges shift to $2\times10^{-14} - 2\times10^{-12}$ mol for NWA 817, $2\times10^{-14} - 4\times10^{-12}$ mol for Zagami, and $3\times10^{-14} - 2\times10^{-12}$ mol for NWA 1068. At 500 μm, the corresponding values are $1\times10^{-12} - 6\times10^{-12}$ mol, $2\times10^{-13} - 7\times10^{-12}$ mol, and $7\times10^{-13} - 3\times10^{-12}$, respectively.

The majority of the $^{40}$Ar content falls below the current mass spectrometer's background level at 50 μm, while most data at 250 μm are above this threshold. An increasing laser diameter leads to greater amounts of extracted $^{40}$Ar, but narrows the K data range. Additionally, if the laser beam profile is Gaussian rather than cylindrical, the ablation volume would be approximately 60% of the cylindrical model. Nevertheless, a 250 μm laser spot is expected to extract an order of magnitude more measurable $^{40}$Ar than the blank level, even after the 40% reduction.

Given these findings, it is expected that a laser spot size of approximately 250 μm will produce $^{40}$Ar data above the detection limit and yield significant K concentration ranges. Note that Fig. 6 is predicted based on a 4000 Ma formation age, while actual ages for NWA 817, Zagami, and NWA 1068 are 1300 Ma, 170 Ma, and 185 Ma, respectively (Mathew et al., 2003; Borg et al., 2005; Shih et al., 2003). Rb–Sr and Sm–Nd ages were assigned to Zagami and NWA 1068, respectively. Since K–Ar isochron ages obtained with the LIBS–MS method should be able to remove the effects of trapped Ar, measured ages are likely close to these ages. Figure 7 shows the calculations adjusted to the actual formation ages of the respective meteorites. The younger formation ages led to lower radiogenic $^{40}$Ar amounts. Consequently, the $^{40}$Ar extracted at 250 μm laser spot with Zagami could potentially fall below the detection limit.

(a) NWA 817

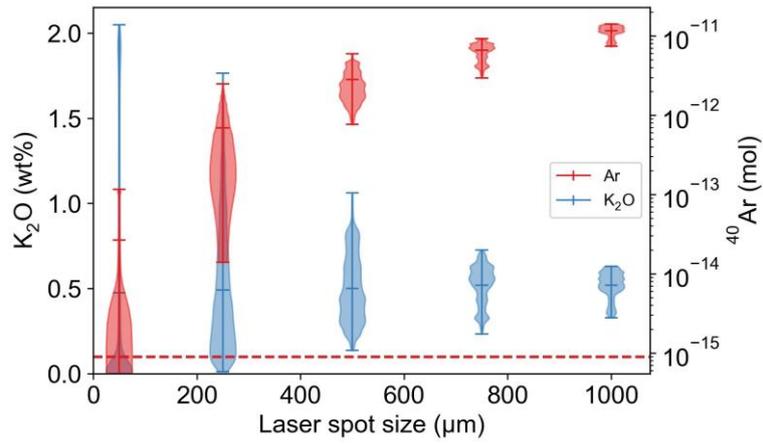

(b) Zagami

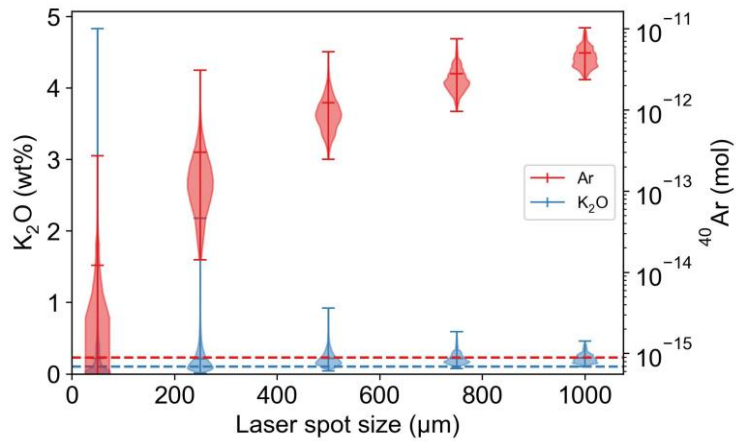

(c) NWA 1068

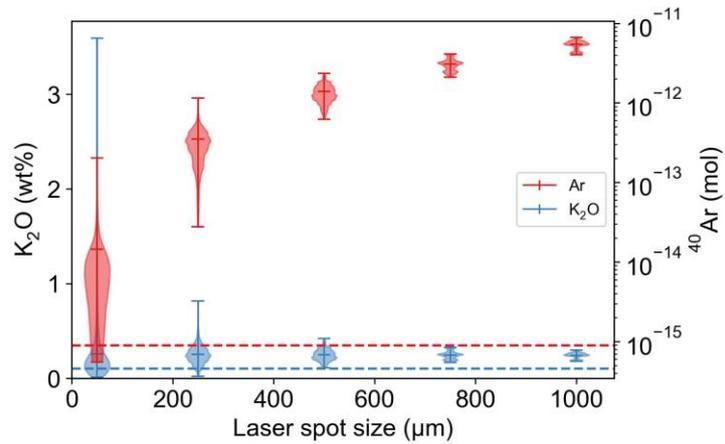

**Figure 6.** Theoretical correlations between laser spot size and measurable Ar and K concentrations if the meteorites (a) NWA 817, (b) Zagami, (c) NWA 1068 were 4000 Ma. Red and blue dashed lines represent the QMS background and the K quantification limit using LIBS (Cho and Cohen, 2018).

(a) NWA 817

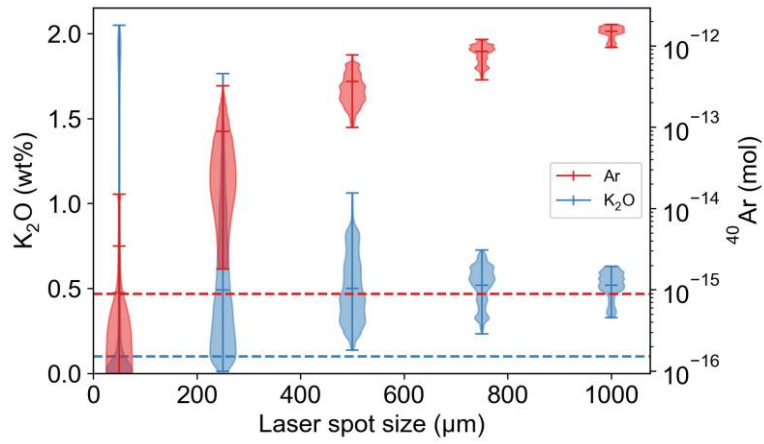

(b) Zagami

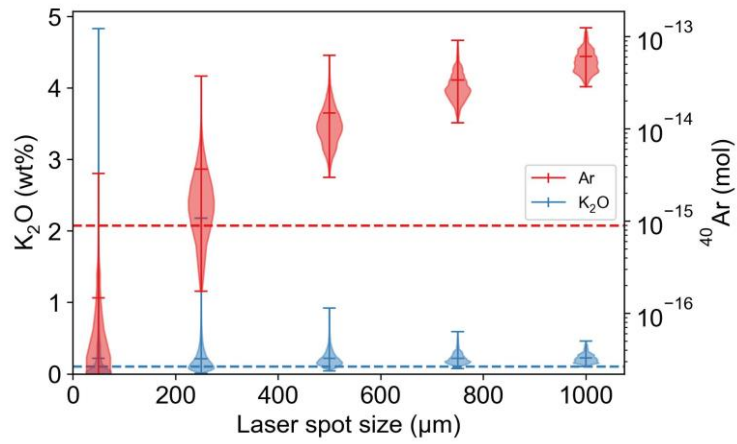

(c) NWA 1068

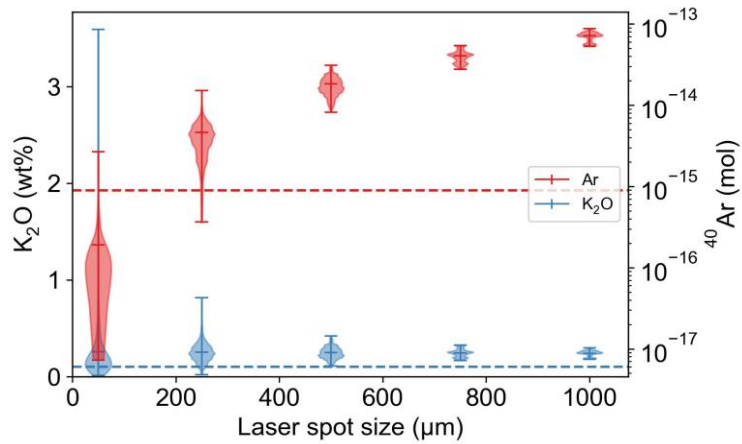

**Figure 7.** Theoretical correlations between laser spot size and measurable Ar and K concentrations for each meteorite based on its actual formation age: (a) NWA 817 (1300 Ma), (b) Zagami (170 Ma), and (c) NWA 1068 (185 Ma).

# Isochron simulation

This section explores the impact of varying five key parameters on the accuracy and precision of isochron dating: mineral composition (meteorite type), laser spot size, the number of isochron data points, formation age, and measurement accuracy for K and Ar concentrations.

*K concentration range*

Initially, we investigate the correlation between the range of K concentration and the precision of isochron dating. The K concentration range represents data point dispersion on the isochron, which directly affects its precision. A narrow range compromises precision of the isochron slope, while a broader range leads to better precision. Figure 8 displays simulated isochrons for the NWA 817 mineral composition with a formation age of 4000 Ma, 15 data points, and measurement errors for K and Ar of 20% each. These isochrons are shown for K concentration ranges of 2, 4, and 8, giving isochron ages of 4843 ± 991 Ma, 3483 ± 640 Ma and 3806 ± 454 Ma, respectively.

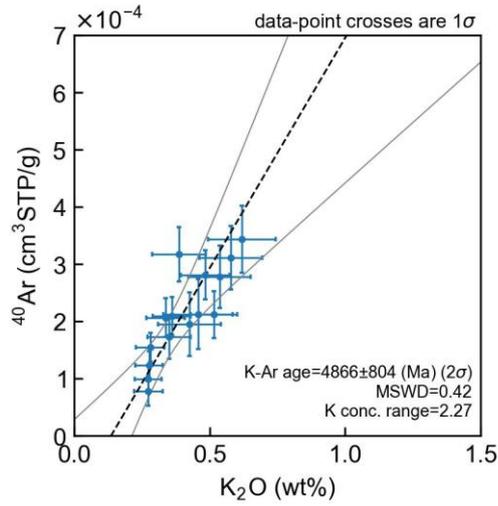

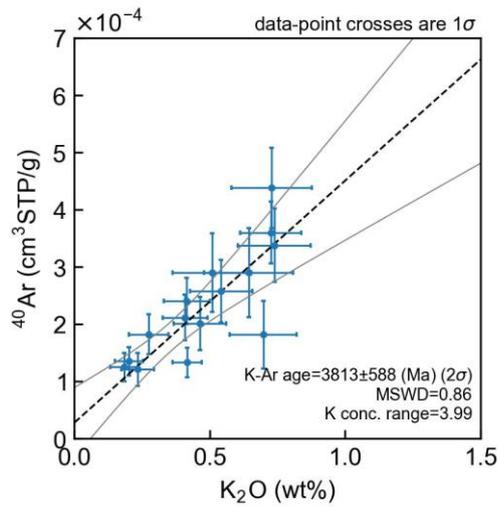

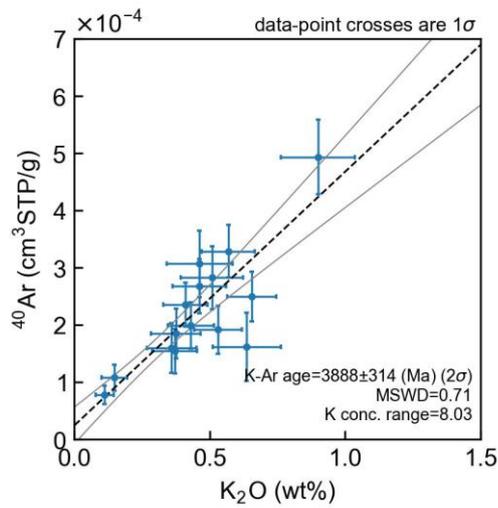

**Figure 8.** Examples of generated isochrons with K concentration range approximately 2, 4, and 8. Potassium map of NWA 817 was used with the parameters ($K_{err}$, $Ar_{err}$, spot, age) = (20 %, 20 %, 15 4000 Ma). Dashed lines represent the best fit isochrons. Gray curves indicate the 95% confidence

interval of each isochron.

Figure 9 shows the relationship between K concentration range and percentage age error based on 100 isochron replicates under different assumptions. They include mineral composition (Fig. 9a), laser spot size (Fig. 9b), and data point number (Fig. 9c). As anticipated, broader K concentration ranges enhanced dating precision in every figure.

For mineral composition, the range of isochron data is affected by mineral size and K concentration. Figure 9a illustrates the precision of isochron ages against the K concentration range of isochrons for each meteorite type, considering a laser spot diameter of 500 μm, a 20% error in K and Ar concentration measurements, 15 data points, and a formation age of 4000 Ma. Each point represents an isochron in each trial, with 100 trials plotted for each meteorite type. NWA 817 and Zagami exhibit similar trends, achieving wide K concentration ranges and high age precision. NWA 817 displays a K concentration range of 3–10 with a dating precision of 10–25%, while Zagami exhibits a K concentration range of 3–13 with a precision of 10–25%. Conversely, NWA 1068 demonstrates a K concentration range of 2–6, with some instances of age precision exceeding 50%. This is because NWA 1068 has a smaller mineral size than the other two, resulting in a narrower K concentration range.

For a given laser spot diameter, larger mineral sizes yielded wider K concentration ranges. Figure 9b presents K concentration ranges versus age precision for four different laser spot diameters (250, 500, 750, and 1000 μm). Each isochron was derived using the mineral distribution of NWA 817, assuming a 20% error in K and Ar concentration measurements, 15 data points, and a formation age of 4000 Ma. At a 1000 μm spot diameter, the K concentration range is 1.5–5, yielding an age precision of 15% or larger, occasionally exceeding 100%. With a 750 μm diameter, the K concentration range spans 1.5–6, resulting in an age precision range of 15–60%. For 250 μm, K concentration range were higher than 4, frequently surpassing 6, with age precision 10–20%.

Figure 9c shows the K concentration range against dating precision for isochrons with spot numbers of 5, 10, 15, and 20. Each isochron is derived from the NWA 817 mineral distribution, considering a laser spot size of 500 μm, a 20% error in K and Ar concentration measurements, and a formation age of 4000 Ma. While the number of data points alone does not significantly widen the range of isochrons, it contributes to the likelihood of capturing a broader range of concentrations. For instance, when the numbers of data points are 5, 10, 15, and 20, the K concentration ranges span 1–8, 1.5–9, 2–12, and 2–10, respectively. Higher data point numbers result in relatively better dating precision, around 15–20%, even with smaller K concentration ranges. Conversely, with only 5 data points, a dating precision of about 60% is observed even with a K concentration range of approximately 6. This discrepancy is due to the increased influence of a single point deviation from the linear trend with fewer data points.

Our parameter studies indicate that the probabilistic age precision variation diminishes when the K concentration range exceeds 6 (Fig. 9). Thus, one should aim for K concentration range higher than 6 to achieve precise isochron measurements.

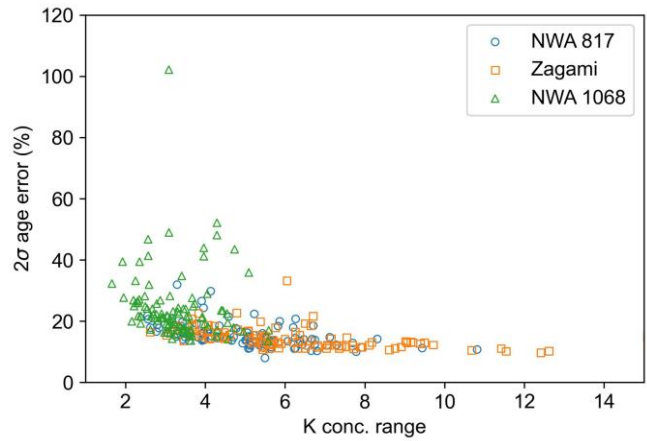

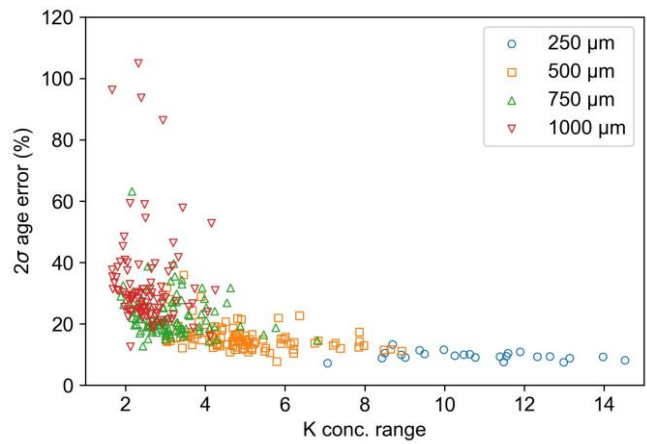

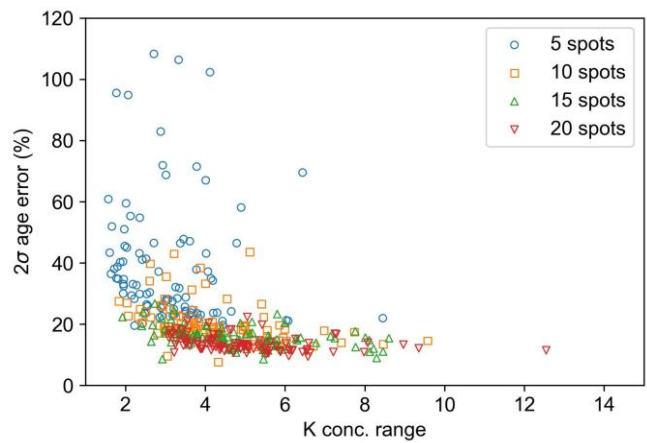

**Figure 9.** Relationship between the range of K concentrations and precision of the ages. (a) For different mineral compositions (meteorite types). Calculations were made under the following conditions: laser spot diameter 500 μm, ($K_{err}$, $Ar_{err}$) = (20%, 20%), formation age 4000 Ma, and 15 data points. (b) For different spot diameters. Mineral composition is of NWA 817, ($K_{err}$, $Ar_{err}$) = (20%, 20%), formation age 4000 Ma, and 15 data points. (c) For different spot numbers. Calculated with a laser spot diameter of 500 μm, mineral composition of NWA 817, ($K_{err}$, $Ar_{err}$)=(20%, 20%), and formation age of 4000 Ma.

*Effect of laser spot size*

This subsection examines the critical role of laser spot diameter in obtaining precise dating results. Figure 10 shows the precision for each meteorite type at different laser spot sizes using box-and-whisker plots. In the box-and-whisker plot, the lower limit of the whiskers represents the first quartile minus 1.5 times the interquartile range, and the upper limit represents the third quartile plus 1.5 times the interquartile range. Values outside these limits are depicted as outliers. These plots reveal how diameter affects the dating precision by influencing the range of measured K concentration (also see Figure 5). The average dating precisions for NWA 817, Zagami, and NWA 1068, excluding outliers, were as follows: 7.8%, 7.7%, and 8.3%, respectively, at a laser spot size of 50 μm; 8.0%, 9.7%, and 14% at 250 μm; 15%, 13%, and 23% at 500 μm; 22%, 17%, and 31% at 750 μm; and 30%, 20%, and 37% at 1000 μm.

      NWA 817 and Zagami exhibit similar average precisions up to a spot diameter of 500 μm. This resemblance is due to their comparable crystal sizes, which enable a sufficient range of K concentration for isochron dating. However, enlarging the spot diameter beyond 500 μm restricts the attainable K concentration range, thereby reducing dating precision. Nevertheless, Zagami offers better dating precision than NWA 817 due to its mesostasis with high local K concentrations.

      NWA 1068's smaller crystal sizes affect its dating precision. At a 50 μm spot diameter, where the crystal size is compatible with the laser spot, the dating precision closely matches that of the other two meteorites. However, at diameters of 250 μm or above, dating precision declines as the measured K concentration range becomes more restricted than that of the other meteorites. This occurs because the laser diameter surpasses the crystal size, leading to a narrower K concentration range.

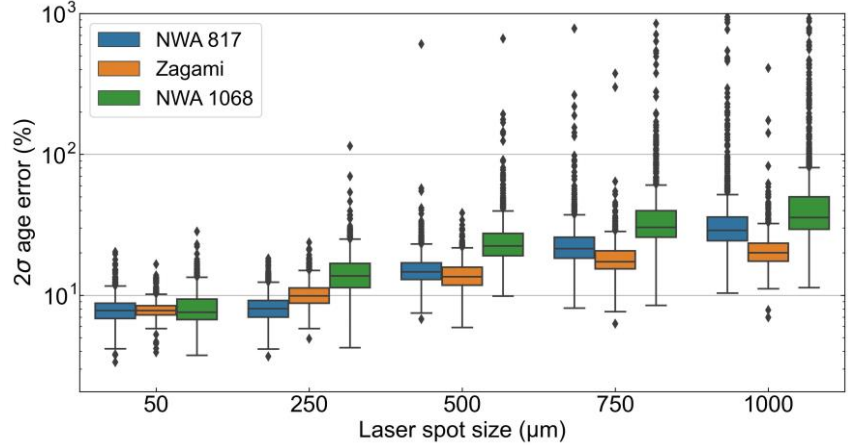

**Figure 10.** Relationship between spot size and dating precision for different mineral compositions. Setting: ($K_{err}$, $Ar_{err}$) = (20%, 20%), age 4000 Ma, 15 spots, 1000 isochron replicates.

*Effect of number of spots*

This subsection addresses the impact of varying the number of laser spots on the precision of isochron dating, considering both the benefits and practical limitations such as time and power constraints on Mars. Increasing the number of data points is generally advantageous for accurate isochron fitting. When rocks are randomly irradiated with a laser for an analysis, increased laser spot numbers would also enhance the chance to sample minor K-rich phases, thereby increasing the K concentration range. Figure 11a visualizes how the precision of isochron dating varies with the number of isochron data points. Statistical analyses were conducted for 1000 isochron replicates with 5 to 50 data points. The average age determination precisions for NWA 817, Zagami, and NWA 1068, excluding outliers, were as follows: 35%, 31%, and 53% for 5 spots; 19%, 17%, and 30% for 10 spots; 15%, 13%, and 23% for 15 spots; and 13%, 11%, and 19% for 20 spots, respectively (Fig. 11a).

To estimate the optimum number of data points, an improvement rate formula is introduced:

$$\text{Improvement rate [\%]} = \{(\Delta t_{i-1} - \Delta t_i)/\Delta t_{i-1}\} \times 100.$$

Here, $\Delta t_i$ represents the average age precision (%) obtained from 1,000 replicate calculations for each spot number condition. "$i$" denotes the number of steps, where $i = 0$ corresponds to 5 spots, $i = 1$ to 10 spots, ..., $i = 9$ to 50 spots.

Increasing the number of spots widens the obtainable K concentration range, improves dating precision, and reduces the likelihood of encountering cases that yield exceptionally poor precision. However, one should stop the measurement at a certain point because of limited in operational resource and mission lifetime. In this context, Fig. 11b shows the diminishing returns in average age precision as the number of laser spots increases. Age precision values drop by approximately 50% when going from 5 to 10 spots, 20–30% from 10 to 15 spots, and another 20% from 15 to 20 points. The rate of improvement became saturated, dropping to less than 10%, when the number of data points exceeded 25. Based on these findings, approximately 15–20 laser spots appear to be sufficient for reliable in-situ K–Ar isochron dating when laser shots are randomly placed.

(a)

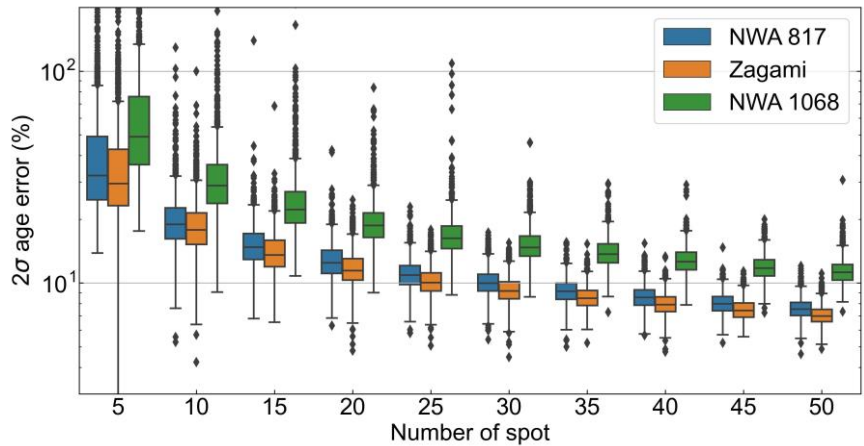

(b)

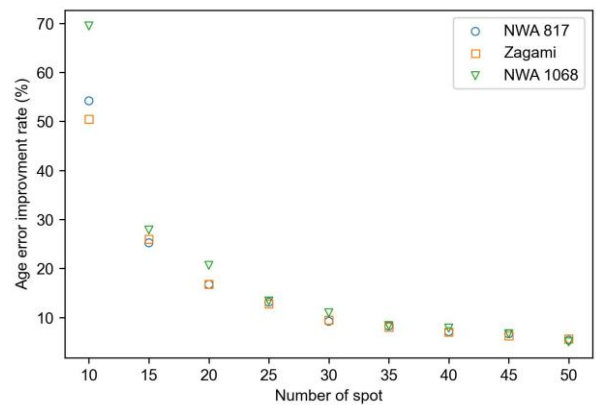

**Figure 11.** (a) Relationship between the number of laser spots and the precision of age determination. (b) Improvement rate in dating precision when varying the number of data points. Setting conditions: ($K_{err}$, $Ar_{err}$)=(20%, 20%), formation age 4000 Ma, 1000 trials, spot diameter = 500 μm, number of spots varied from 5 to 50 in increments of 5.

*Effect of formation age*

This subsection examines the impact of formation age on dating precision. Figure 12 presents the data obtained from 1000 fitting trials. All replicates used a K and Ar concentration measurement error of 20%, 15 data points, and a 500 μm spot diameter. NWA 817 and Zagami exhibit comparable dating precision due to their similar mineralogical characteristics; the size, K concentration, and modal abundance of K-rich and K-poor phases are comparable to each other (Fig. 4b, 4d). For rocks with a mineralogical characteristic similar to NWA 817, Zagami, and NWA 1068, average precisions for in situ K–Ar dating at different geologic ages are as follows: 15%, 13%, and 23% for 4000 Ma; 18%, 16%, and 27% for 3000 Ma; 22%, 20%, and 34% for 2000 Ma; and 28%, 26%, and 43% for 1000 Ma, respectively. Relative precision improves as the formation age increases, consistent with the experimental results by Cho et al. (2016).

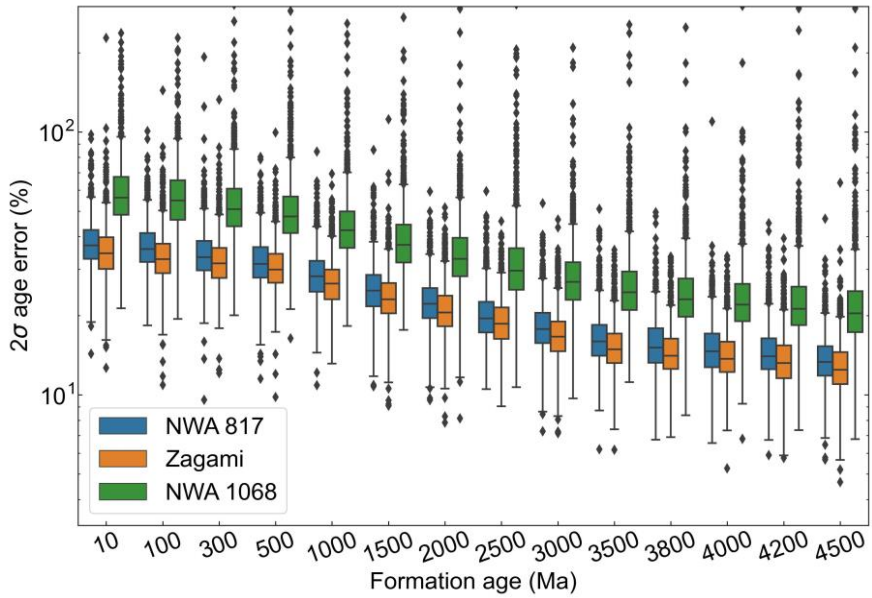

**Figure 12.** Relationship between an age and dating precision. ($K_{err}$, $Ar_{err}$)=(20%, 20%), 15 spots, 1000 trials, spot diameter = 500 μm.

*Effect of K and Ar measurements errors*

Measurement errors in K and $^{40}$Ar concentrations have direct impact to precision of age estimation. Fig. 13 illustrates how varying these errors affects the age precision. Each pixel in the figure represents the average precision from 100 isochron replicates, assuming specific K and Ar error values at each coordinate ($x$, $y$). Our model parameters were 15 measurement spots, a 500 μm spot diameter, and a formation age of 4000 Ma.

With K and Ar concentration errors of 20%, the 2σ age precision for NWA 817, Zagami, and NWA 1068 was ±600 Myr, ±570 Myr, and ±1400 Myr, respectively. Reducing the errors to 10% improved the age precision to ±320 Myr for NWA 817, ±290 Myr for Zagami, and ±730 Myr for NWA 1068. An even further reduction in errors to 5% enhances the age precision to ±160 Myr, ±150 Myr and ±380 Myr for NWA 817, Zagami and NWA 1068, respectively. NWA 817 and Zagami exhibit comparable age precisions, largely because their mineral sizes align well with the 500 μm laser spot diameter, leading to similar K concentration ranges. In contrast, NWA 1068 shows lower precision due to the narrower K concentration range when using the 500 μm spot diameter.

Our results above indicate that, for volcanic rocks with NWA 817-like and Zagami-like mineral compositions with a formation age 4000 Ma, using a 500 μm laser spot diameter can yield a 2σ age precision of 200 Ma, provided the K and Ar concentration measurement errors are below 7%. Achieving comparable precision for NWA 1068, which has a smaller crystal size than NWA 817 and Zagami, was found more challenging, requiring K and Ar measurement accuracy of approximately 3% if the laser spot diameter is 500 μm.

(a) NWA 817

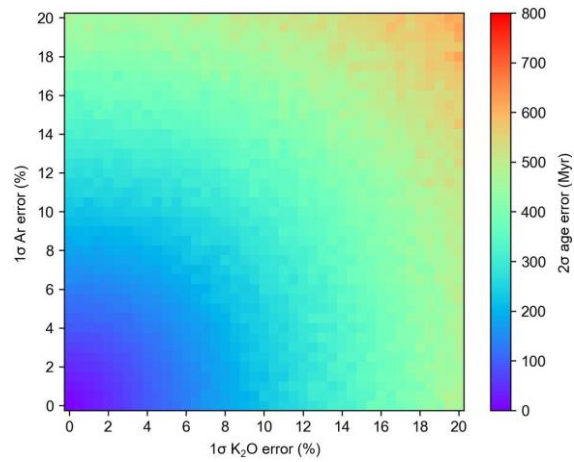

(b) Zagami

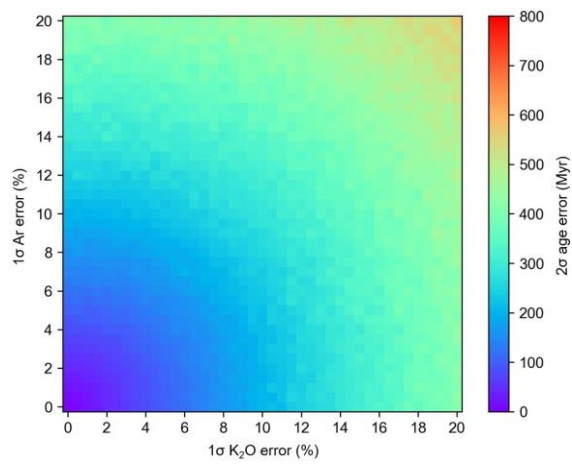

(c) NWA 1068

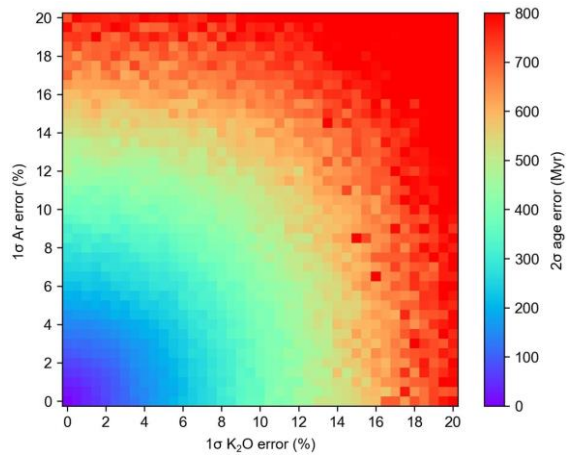

**Figure 13.** Age determination errors for various K and Ar measurement errors. (a) NWA 817 (b) Zagami (c) NWA 1068. Calculation parameters: formation age 4000 Ma, spot diameter 500 μm, number of spots 15. ($K_{err}$, $Ar_{err}$) were varied in 0.5% increments. 100 replicates for each parameter set.

*Dating accuracy*

While the previous sections focused on the precision of age determination, precision alone does not guarantee accuracy. In this section, we investigate how closely the calculated ages align with the true ages, defining dating accuracy as the deviation between these ages. Relative accuracy is calculated as the ratio of the calculated age to the true age, initially set as a model parameter. A ratio of 1 indicates perfect agreement between the two ages. This deviation should be comparable or smaller than the dating precision.

Figure 14a illustrates the impact of the number of laser spots on relative accuracy, assuming a spot diameter of 500 μm, K and Ar concentrations measurement errors of 20%, and a formation age (true age) of 4000 Ma. The age accuracy improves as the number of spots increases. With 15 laser spots, better than 10% accuracy was attainable with a 50% probability for any examined meteorite.

Figure 14b explores age accuracy across different formation ages, using 15 spots and a spot diameter of 500 μm, K and Ar concentrations measurement errors of 20%. As expected, accuracy improved with increasing sample age. Maximum deviations within the interquartile range remained under 10%. The mean value across all age ranges approximates 1, indicating that calculated ages were neither underestimated nor overestimated. For samples similar to NWA 817 or Zagami, approximately 50% of the data points fell within the following ranges: 1 ± 0.25 for an age of 1000 Ma, 1 ± 0.10 for ages of 2000 and 3000 Ma, and 1 ± 0.05 for an age of 4000 Ma. These results indicate that the average accuracy of age determination exhibits smaller variability compared to the precision calculated with the same parameter setting: the average precision of age determination was 26% for 1000 Ma, 20% for 2000 Ma, 16% for 3000 Ma, and 13% for 4000 Ma for Zagami, which has the lowest precision (Fig. 12). This comparison between accuracy and precision in dating suggests that the true value of a sample age is encompassed within the error range specified by its precision.

Figure 14c shows the average age values from 100 replicated isochrons, calculated under different K and Ar concentration errors. A formation age of 4000 Ma was assumed. We found that in the isochron fitting with the York method, larger errors in K concentration (*x*-error) lead to overestimation of age values. Conversely, larger errors in Ar concentration tend to result in underestimation of the age. This tendency is particularly pronounced when either error is significantly larger than the other. The deviation from the true values in the average of 100 trials is less than 5%. Recognizing the possibility of this method-specific bias is essential if varying measurement errors exist between K and Ar concentrations. We also found that the systematic deviation of the calculated age values (Fig. 14c) from the true values is inevitably caused by the weighting algorithm of the York method. In the datasets where the data points are concentrated in a small area, if x-errors are much larger than y-errors, the York method tries to fit the data points in the y-axis direction, resulting in a steeper slope of the fitted line. Conversely, if the error bars in the y-axis direction are larger, the slope of the isochron becomes gentler due to adjustments in the opposite direction (Fig. S2). It should be noted that such systematic errors are likely to occur in actual measurements if the K and Ar measurement errors are not equal, which is sometimes the case.

(a)

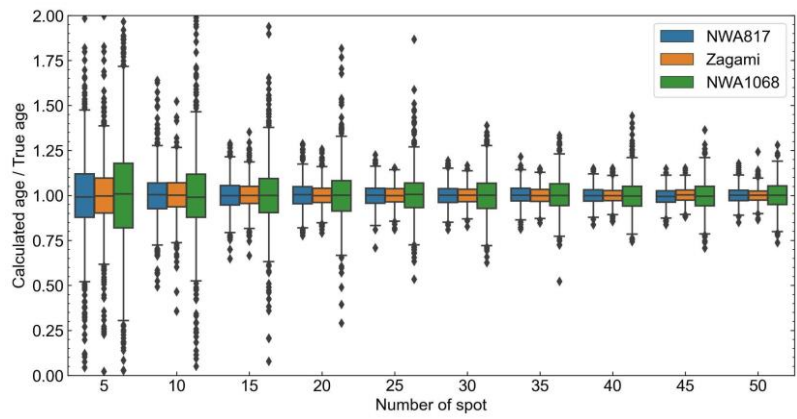

(b)

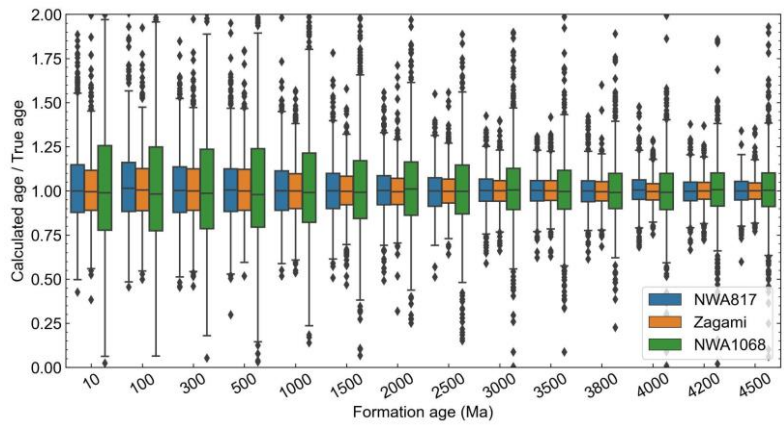

(c)

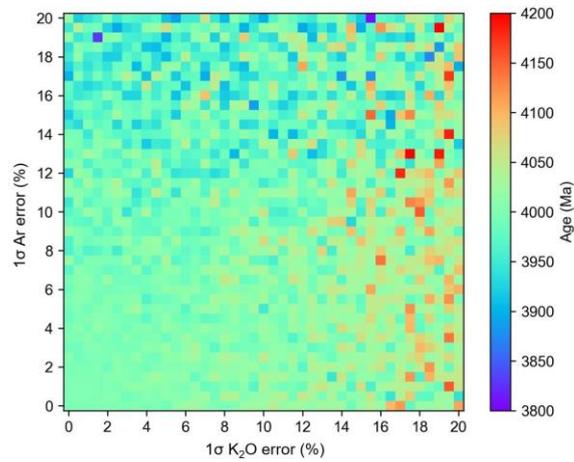

**Figure 14.** (a) Accuracy of age values when the number of spots was varied. The age value of 4000 Ma was assumed. (b) Variation of accuracy for different formation ages. (c) Age values for NWA 1068 when K and Ar concentrations were varied. The age value of 4000 Ma was assumed.

# DISCUSSION

## Requirements for high precision dating

Our numerical experiments revealed that the laser spot diameter of 500 μm is capable of exceeding a K concentration range of 6 for rocks resembling the NWA 817 nakhlite in terms of mineralogy. The mineralogy of NWA 817, characterized by large crystals, allows for a wide range of K concentrations when using the laser diameter 500 μm or smaller. For Zagami, which has a crystal size similar to or slightly larger than NWA 817, the distribution of K concentrations is similarly broad. In contrast, NWA 1068, with its smaller crystal size, exhibits a narrower K concentration distribution under the same laser conditions. Reducing the laser diameter further expands the measurable K concentration range due to the presence of localized K-rich zones. Our calculations suggest that a 5% precision in dating can be attained if the K concentration range is greater than 6 and the errors for K and Ar concentrations are 20%. Nonetheless, the probability of achieving this precision drops to less than 0.1%. Figure 15 presents dating precision under different instrument settings with achievable K concentration ranges. The formation age was assumed to be 4000 Ma, and 1000 Monte-Carlo runs were made under the same conditions. The blue dots correspond to the typical LIBS–MS experiment settings, which include a 500 μm laser diameter and 20% measurement errors for both K and Ar concentrations (Cohen et al., 2014; Cho & Cohen, 2018; Cattani et al., 2023). Our analyses indicate that achieving a dating precision of 5% is challenging with these conditions. While reducing the measurement errors improves a dating precision, as shown by the orange and green dots, the large laser diameter prevents the attainment of a sufficient range of K concentrations. This result highlights the need for smaller laser spot diameters as well as higher measurement accuracies in both elements to improve the dating precision.

Based on our analysis of the three Martian meteorites, a laser spot diameter of 250 μm and a measurement accuracy of 10% for K and Ar concentrations could achieve an age determination precision of 200 Myr (2σ) for a 4000 Ma sample (dashed line in Fig. 15), as indicated by the red dots (Fig. 15). This capability meets specifications outlined in NASA technology roadmaps (2015) and the NASA Decadal survey (National Aeronautics and Space Administration, 2015; National Research Council, 2011). While many points in these plots did not achieve the desired 5% age precision, particularly when the K concentration range was smaller than 6, targeted laser irradiation would allow for obtaining such a wide range of K concentrations. By mapping K concentrations through preliminary laser shots across the sample surface one can strategically select irradiation points to achieve a sufficiently broad K concentration range. Our results indicate that the probability of achieving a dating precision of <5% is 83%, 54%, and 15% for NWA 817, Zagami, and NWA 1068, respectively, with a 250 μm laser irradiation onto random locations and 10% measurement accuracies for K and Ar concentrations. Conducting 15 or more spot analyses on different mineral phases is effective in achieving a 5% dating precision, especially for rocks with mineralogy similar to Zagami or NWA 817.

Existing instrumentation on Mars missions supports the feasibility of our recommendations for laser diameter and measurement error. For instance, the Mars Surface Composition Detector (MarSCoDe) on China's Tianwen-1 mission achieved a laser diameter of less than 200 μm at a target distance of approximately 2 m (Xu et al., 2021). Thus, a laser diameter of 250 μm is an achievable target. Regarding Ar concentration measurements (mol/g), a 10% measurement accuracy has already been realized in previous studies (e.g., Cattani et al., 2023). However, achieving a 10% measurement accuracy for K concentration remains a challenge. Improving the accuracy would require expanding calibration samples with low K concentrations (< 1 wt%).

Another approach for enhancing dating precision is to select rocks with larger mineral sizes and higher K contents. The Perseverance rover discovered an igneous rock that is believed to have formed through the deposition of olivine crystals in magma (Liu et al., 2022). This rock exhibits a crystal size ranging from 0.5 to 3.0 mm, exceeding both the crystal sizes analyzed in this study and the laser diameter typically used for LIBS–MS instruments. Furthermore, the compositional measurements with the Planetary Instrument for X-ray Lithochemistry (PIXL) instrument revealed that this rock also has a variation in K concentration in each mineral phase. The modal abundance of igneous rocks to be 60–70% for olivine, 5–18% for augite, and 7–13% for mesostasis. The measured $K_2O$ concentrations using XRF were 0.00 wt% for olivine, 0.00 wt% for pyroxene, and 0.32–2.18 wt% for mesostasis (Liu et al., 2022). These levels of K concentration are comparable to those observed in NWA 817 in this study. Consequently, the substantial crystal sizes in these Martian rocks would enable separate spot analyses of individual minerals, potentially yielding an isochron with a K concentration range exceeding 6. Such an approach would be capable of achieving a dating precision (2σ) of 200 Myr for samples with the age of ~4000 Ma.

(a) NWA 817

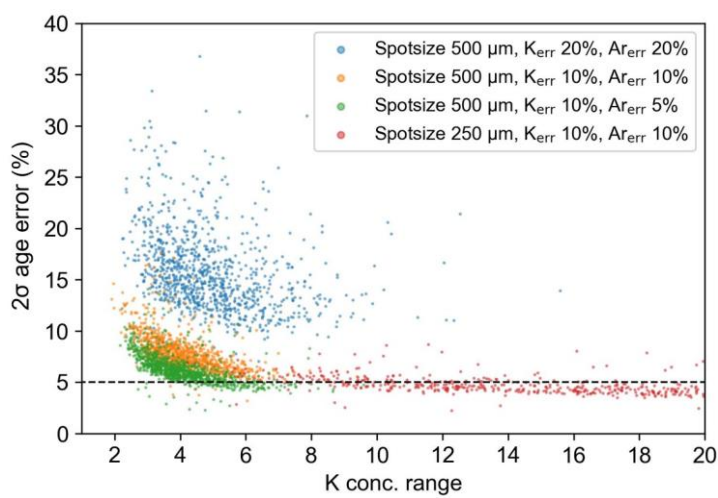

(b) Zagami

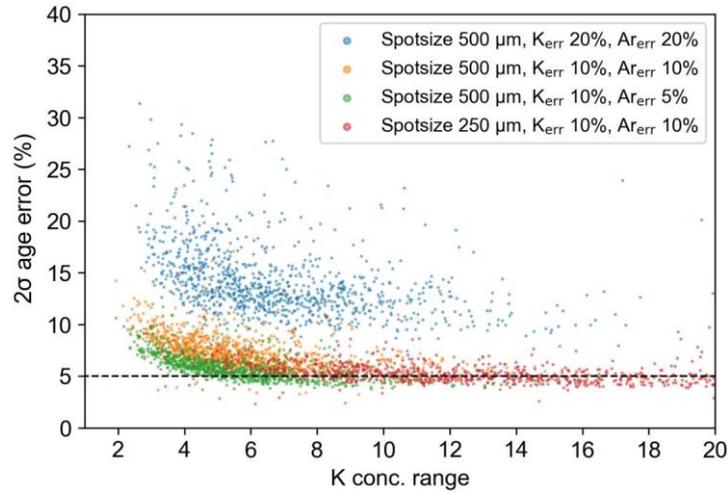

(c) NWA 1068

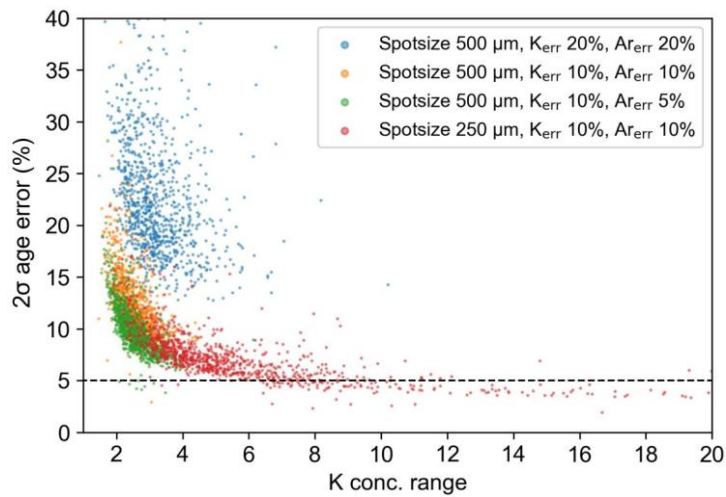

**Figure 15.** Statistical distribution of age measurement precision under various parameters. Samples are (a) NWA 817 (b) Zagami, and (c) NWA 1068. Each point corresponds to one isochron simulation. Dashed lines indicate 5% error level.

## Implications for geochronology measurements on Mars

The dating precision achieved in this study has the potential to resolve a range of scientific questions. One key application is calibrating crater chronology through radiometric ages. The absolute age of a sample originated from a geologic unit with well-defined crater density enables the refinement of crater chronology models for Mars (Werner and Ivanov, 2015).

Werner (2019) attempted such calibration on Mars using the K–Ar model ages measured by the Curiosity rover, but the age and its error of 4.21 ± 0.35 Ga was inadequate for constraining the crater age models. If multiple rocks with formation ages of 3 Ga and 4 Ga could be dated with a 5% precision at different locations on Mars, as demonstrated in this study, it would be possible to constrain the crater age models that is consistent with the radiometric age data.

One of the critical uncertainties in Martian crater age models pertains to the Noachian–Hesperian boundary, with age estimates varying from 3.9 to 3.6 Ga (Cohen et al., 2021). This boundary is thought to correspond with Mars' transition from a warm and wet climate to its present cold and dry state (e.g., Wordsworth, 2016). Dating samples from this geological period could determine the absolute age of this climatic transition and refine existing crater chronology models, thereby revealing the timeline of loss of habitability on Mars. In situ analyses of rocks similar to NWA 817 or Zagami, using a spot diameter of 500 μm, are expected to yield an age precision of 600 Myr (2σ), a level of precision insufficient for distinguishing age boundaries. However, employing a laser spot diameter of 250 μm for these analyses could improve age precision to 200 Myr (2σ), thereby reducing the uncertainty of the age boundary.

Analyzing Amazonian rocks on Mars, which are believed to have ages of 1–3 Gyr, represents another important application of in situ geochronology. This measurement is key for calibrating crater ages and reducing uncertainties in the impact flux of the recent past (Cohen et al., 2021). The recent lunar sample returned by the Chang'e-5 lunar lander, dated at 2.0 Ga (Yue et al., 2022), provides a critical anchor within the 1–3 Ga period for the lunar crater chronology model. However, measuring Amazonian rocks on Mars is essential to verify the applicability of the lunar chronology model to Martian conditions. In this context, our study revealed that with a laser spot diameter of 250 μm and a 10% measurement accuracy for both K and Ar, it is possible to achieve significant dating precision for rocks similar to NWA 817. Figure 16 illustrates the precision of isochron ages across various formation ages, employing a proposed laser spot diameter of 250 μm, K and Ar concentration measurement accuracy of 10%, and based on 15 data points. Specifically, the expected 2σ dating precisions for rocks of 1, 2, and 3 Ga are 80 Myr, 125 Myr, and 150 Myr, respectively (Fig. 16).

In addition to the crater chronology calibration, in situ radiometric dating of igneous rocks derived from volcanic eruptions and lava flows, indicates the timing of volcanic activity on Mars. Combined with mineral composition information, these ages provide constraints on eruption magnitudes and mineral crystallization processes. Furthermore, volcanic eruptions have a significant impact on Martian atmospheric pressure and composition. For example, volcanic outgassing between 3.5 and 2 Ga likely produced sufficient $CO_2$ and $H_2O$ to create an atmospheric pressure of 1 bar (Grott et al., 2011). This activity also contributed methane and other greenhouse gases to the atmosphere. Determining the timing of such volcanic episodes helps establish the timing and scale of major gas emissions. A comprehensive understanding of Martian volcanic history could elucidate the planet's atmospheric evolution (Mouginis-Mark et al., 2022). In situ dating with our method can achieve an error of 125–160 Myr for Martian rocks of 2–3.5 Ga. This level of precision is achievable with a 250 μm laser spot diameter and a 10% measurement accuracy for both K and Ar, specifically for rocks analogous to NWA 817. Such precision in dating is useful in constraining the timing of large-scale volcanic activities on Mars by measuring the age of multiple volcanic zones and lava flows such as Tharsis region and Elysium Planitia (Broquet and Andrew-Hanna, 2023).

Radiometric dating of sedimentary rocks for ascertaining the timing of fluvial activities is less straightforward due to challenges in interpreting the resultant ages, which do not necessarily reflect the time of deposition. On Earth, the conventional methodology involves investigating the stratigraphic sequences and deriving the absolute age of the sedimentary layers by dating adjacent layers of volcanic products. A similar strategy would be applicable for in situ K–Ar dating on Mars as well. Curiosity has conducted K–Ar dating of sedimentary rocks on Mars (Martin et al., 2017), primarily composed of plagioclase and jarosite. The bulk K–Ar age was 2.57 Ga. However, by leveraging the differential Ar retention between plagioclase and jarosite and extracting Ar through two stages of heating at different temperatures, K–Ar model ages of 4.07 Ga for plagioclase and 2.12 Ga for jarosite were determined (Martin et al., 2017). This suggests that K–Ar ages of sedimentary rocks may represent a mix of ages from detrital igneous components and fluid-associated diagenetic/authigenic phases. If these phases are resolved through laser ablation, it would be possible to derive two distinct isochron ages from such rocks.

Note that this study did not account for the resetting of K–Ar ages or the occurrence of different minerals of different ages in a single rock. If a K–Ar age is completely reset, the measured age would reflect the age of the heating event (e.g., impacts) rather than the rock's formation age. In such cases, the K–Ar age corresponds the absolute age of crater formation, valuable data for calibrating crater chronology models as well as local geology of the explored region. In contrast, partial resetting of the K–Ar system could be indicated by non-zero intercepts of the isochron (Cosca et al. 1998). If a rock contains minerals with disparate age values, determining the formation age of the rock becomes complex, leading to isochron data points that diverge from a linear pattern. To resolve this issue, one could reduce the laser spot diameter to be smaller than the mineral grain size, allowing for discrete mineral age measurements and further insights into the rock's formation history. Achieving such measurements requires a laser diameter smaller than the size of a dominant mineral and a protocol for identifying rocks that have larger mineral sizes than the laser diameter before dating experiments.

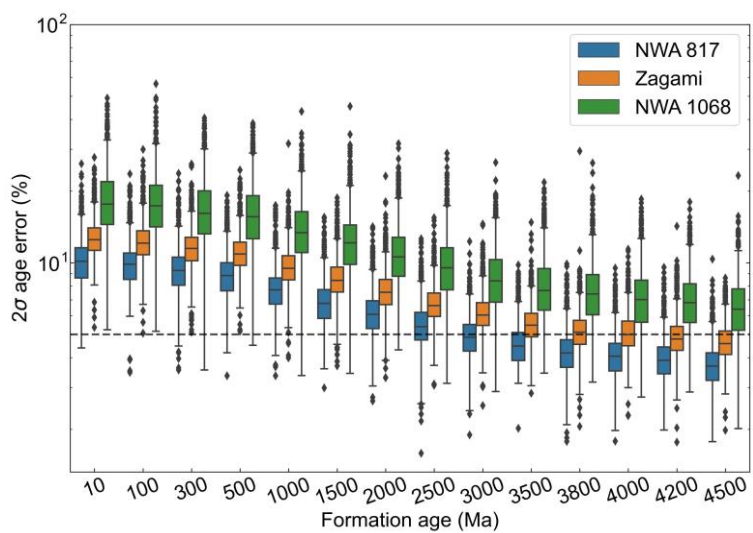

**Figure 16.** Dating precision as a function of age. Instrument conditions were determined from the results of this study to achieve a dating precision of 5%: ($K_{err}$, $Ar_{err}$)=(10%, 10%), 15 spots, spot diameter = 250 μm, 1000 trials. Dashed lines indicate 5% error level.

# CONCLUSIONS

This study uses data on the mineral composition of Martian meteorites, representative of Martian rocks, to simulate the isochrons of a broad range of Martian rocks with formation ages ranging from 10 Ma to 4500 Ma. We employed laser spot analysis with the LIBS–MS method and varied several parameters to obtain quantitative statistics on age precision.

Throughout the analysis, we conducted isochron simulations focusing on five key parameters. These parameters were (1) mineralogical texture derived from actual Martian meteorites, (2) assumed formation ages of the rocks, (3) laser spot diameter, (4) number of measured data points, and (5) measurement error of K and Ar concentrations.

First, regarding rock texture, we observed that a wider range of K concentrations can be obtained on the isochrons if the mineral size exceeds the laser spot diameter, leading to higher dating precision. For example, NWA 1068, primarily composed of olivine grains of 100–300 μm, exhibited K concentration ranges of 0.11–0.42 wt% at 500 μm, and 0.02–0.81 wt% at 250 μm. This observation suggests that even in rocks with low bulk concentrations, areas with significant K concentration allow for obtaining a significant range on the isochron for precise age determination through laser spot analysis.

Second, we quantified improving relative precision of dating as the formation age of rocks increase. For example, with a 250 μm laser spot diameter and 10% measurement errors in K and Ar concentrations, $2\sigma$ dating precisions of 80 Myr, 125 Myr, 150 Myr, and 166 Myr can be achieved for NWA 817-like rocks with formation ages of 1000 Ma (relative precision of 8%), 2000 Ma (6%), 3000 Ma (5%), and 4000 Ma (4%), respectively. These values satisfy the 200 Myr precision required by the NASA Technology Roadmaps (National Aeronautics and Space Administration, 2015).

Third, while employing a smaller laser spot diameter to measure minerals separately can increase the K concentration range on isochrons, this also results in a smaller volume vaporized by the laser, thereby reducing the amount of extracted Ar. This presents a trade-off between the K concentration range and the quantity of extracted Ar. Assuming 4000 Ma rocks, our analysis of three Martian meteorites suggests that a 250 μm laser spot diameter is optimal, simultaneously ensuring a sufficiently wide range of K concentrations and a sufficient amount of Ar extraction above the background level of the mass spectrometer used in published laboratory studies.

Fourth, the number of data points measured was evaluated by assessing improvements in dating accuracy with an increase in data points from 5 to 50, in increments of 5. Our results indicate that improvements of over 10% are attainable up to 15–20 data points, with minor enhancements beyond that number. Given the operational constraints in in situ measurements, measuring 15–20 data points are sufficient for satisfactory precision.

Finally, measurement errors in K and Ar concentrations are reflected in isochron data points' error bars. With K and Ar concentration errors of 20%, laser spot size of 500 μm, and formation age of 4 Ga, the 2σ age precision for NWA 817 was ±600 Myr. Reducing the errors from 20% to 10% improved the age precision to ±320 Myr and an even further reducing the errors to 5% improved the age precision to ±160 Myr. To achieve 200 Myr precision, reducing laser spot diameter to 250 μm would be more effective than trying to achieve 5% measurement errors. In terms of dating accuracy, with K and Ar concentration errors of 20%, laser spot size of 500 μm for Zagami, approximately 50% of the data points fell within the following ranges: 1 ± 0.25 for an age of 1000 Ma, 1 ± 0.10 for ages of 2000 and 3000 Ma, and 1 ± 0.05 for an age of 4000 Ma. These results indicate that the average accuracy of age determination exhibits smaller variability compared to the precision calculated with the same parameter setting. Furthermore, we found that in the isochron fitting with the York method, larger errors in K concentration than in Ar lead to overestimated ages, while larger errors in Ar concentration lead to underestimated ages. This tendency becomes more pronounced when one error is substantially larger than the other.

In summary, our quantitative isochron modelling using mineral textures of actual Martian meteorites revealed that to achieve a dating precision of 200 Myr in 2σ significance, optimal conditions include a laser spot diameter of 250 μm and a measurement uncertainty of 10% for K and Ar concentrations. Achieving a 250 μm laser spot diameter is feasible with existing flight instruments. In addition, 10% Ar measurement error can be achieved with a QMS, technique already used in the Sample Analysis at Mars (SAM). Achieving a 10% measurement error for K concentration would require calibration experiments using standard samples with low K concentrations (< 1 wt%). These findings suggest that in situ dating with 200 Myr (2σ) precision is feasible on the Martian surface once these improvements are implemented.

*Acknowledgements*

The authors thank Mr. Hideto Yoshida, Department of Earth and Planetary Science, the University of Tokyo, for supporting the EPMA analysis. This study was supported by JSPS grant-in-aid JP23H01225. H. Hyuga gratefully acknowledges the support by International Graduate Program for Excellence in Earth-Space Science (IGPEES) from the University of Tokyo.

Treiman, Allan H. 2005. The Nakhlite Meteorites: Augite-Rich Igneous Rocks from Mars. *Geochemistry: Exploration, Environment, Analysis* 65 (3): 203–70.

Treiman, Allan H. 2021. Uninhabitable and Potentially Habitable Environments on Mars: Evidence from Meteorite ALH 84001. *Astrobiology* 21 (8): 940–53.

Udry, Arya, and James Day. 2018. 1.34 Billion-Year-Old Magmatism on Mars Evaluated from the Co-Genetic Nakhlite and Chassignite Meteorites. *Geochimica et Cosmochimica Acta* 238 (10): 292–315.

Vasconcelos, P. M., K. A. Farley, C. A. Malespin, P. Mahaffy, D. Ming, S. M. McLennan, J. A. Hurowitz, and Melissa S. Rice. 2016. Discordant K-Ar and Young Exposure Dates for the Windjana Sandstone, Kimberley, Gale Crater, Mars. *Journal of Geophysical Research. Planets* 121 (10): 2176–92.

Werner, S. C., and B. A. Ivanov. 2015. 10.10 - Exogenic Dynamics, Cratering, and Surface Ages. In Treatise on Geophysics (Second Edition), edited by Gerald Schubert, 327–65. Oxford: Elsevier.

Werner, Stephanie C. 2019. In Situ Calibration of the Martian Cratering Chronology. *Meteoritics & Planetary Science* 54 (5): 1182–93.

Wordsworth, Robin D. 2016. The Climate of Early Mars. *Annual Review of Earth and Planetary Sciences* 44 (1): 381–408.

Xu, Weiming, Xiangfeng Liu, Zhixin Yan, Luning Li, Zhenqiang Zhang, Yaowu Kuang, Hao Jiang, et al. 2021. The MarSCoDe Instrument Suite on the Mars Rover of China's Tianwen-1 Mission. *Space Science Reviews* 217 (5): 64.

York, Derek, Norman M. Evensen, Margarita López Martínez, and Jonás De Basabe Delgado. 2004. Unified Equations for the Slope, Intercept, and Standard Errors of the Best Straight Line. *American Journal of Physics* 72 (3): 367–75.

Yue, Zongyu, Kaichang Di, Wenhui Wan, Zhaoqin Liu, Sheng Gou, Bin Liu, Man Peng, et al. 2022. Updated Lunar Cratering Chronology Model with the Radiometric Age of Chang'e-5 Samples. *Nature Astronomy* 6 (5): 541–45.

# APPENDIX

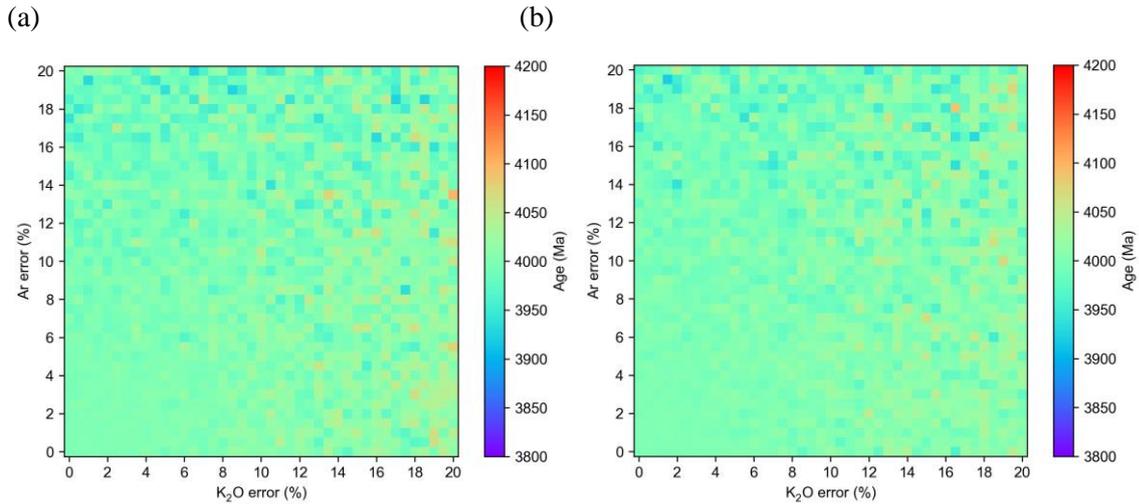

**Figure S1.** Age values for NWA 1068 when potassium and Ar concentrations were varied. The age value of 4000 Ma was assumed.

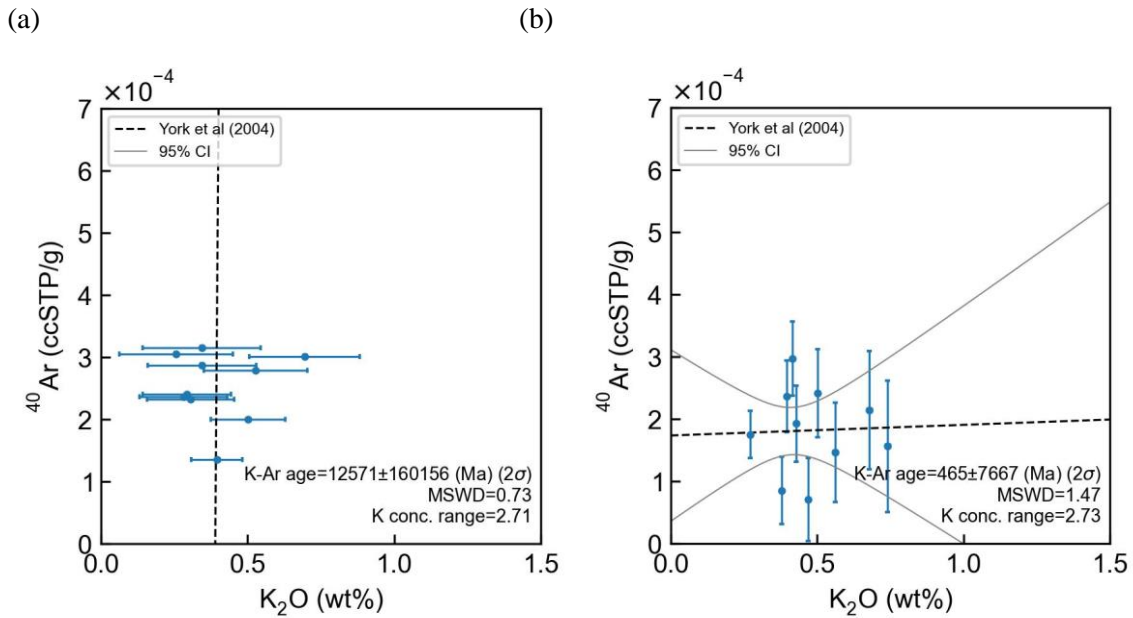

**Figure S2.** Examples of when the range of data points is narrow and isochron age values are significantly overestimated or underestimated when measurement errors are biased. (a) when the error in the K concentration (error in the x-axis direction) is large; (b) when the error in the Ar concentration (error in the y-axis direction) is large.